\documentclass[12pt,aps,prd,superscriptaddress,showpacs,floatfix,nofootinbib]{revtex4-1}
\pdfoutput=1

\usepackage{bmpsize}

\usepackage{hyperref}
\usepackage{color}
\usepackage{graphicx}	

\usepackage{bm}
\usepackage{amsmath}
\usepackage{amsfonts}
\usepackage{eufrak}

\begin{document}

\title{Lattice Simulations of 10d Yang-Mills toroidally compactified to 1d, 2d and 4d}

\author{Masanori Hanada}
\affiliation{Yukawa Institute for Theoretical Physics, Kyoto University,\\
Kitashirakawa Oiwakecho, Sakyo-ku, Kyoto 606-8502, Japan}	
\affiliation{The Hakubi Center for Advanced Research, Kyoto University,\\
Yoshida Ushinomiyacho, Sakyo-ku, Kyoto 606-8501, Japan}
\affiliation{Stanford Institute for Theoretical Physics,
Stanford University, Stanford, CA 94305, USA}
\author{Paul Romatschke}
\affiliation{Department of Physics, University of Colorado, Boulder, Colorado 80309, USA}
\affiliation{Center for Theory of Quantum Matter, University of Colorado, Boulder, Colorado 80309, USA}

\begin{abstract}
Toroidally compactified Yang-Mills theory on the lattice is studied by using the Hybrid Monte Carlo algorithm. When the compact dimensions are small, the theory naturally reduces to Yang-Mills with scalars. We confirm previous analytical and numerical results for pure gauge theory with scalars in $(0+1)$ dimensions and at high temperatures to Super-Yang-Mills in $(1+1)$  dimensions.
In $(1+1)$ dimensions, our simulations confirm the previously conjectured phase diagram. Furthermore, we find evidence for the sequential breaking of the center symmetry in $(1+1)$ dimensions as a function of the volume.
In $(3+1)$ dimensions we present first simulation results for the eigenvalue distribution of the Polyakov and Wilson loops, finding localized, non-uniform and center-symmetric configurations as a function of the lattice coupling.
\end{abstract}

\maketitle
\section{Introduction}
Toroidally compactified Yang-Mills theories on a $d$-dimensional lattice are of interest for several reasons. 
The case of $d=4$ has obvious applications to QCD; sometimes calculations simplify at small volume, 
and one can have some hope to learn lessons about the large-volume theory; see e.g. Refs.~\cite{Eguchi:1982nm,Luscher:1982ma,Unsal:2008ch,Narayanan:2003fc}. 
Another important application is the gauge-Higgs unification scenario; 
just as a four-dimensional vector field is obtained from the five-dimensional metric via the Kaluza-Klein mechanism \cite{Kaluza:1921tu,Klein:1926tv}, 
scalars in four dimensions can be obtained from five-dimensional vectors via compactification; see e.g. Refs.~\cite{Manton:1979kb,Hosotani:1983xw} for previous work on this subject.
Certain phenomenological models with large extra dimensions, such as the universal extra dimensions \cite{Appelquist:2000nn}, add further motivations. 
Lattice studies along these directions can be found e.g. in Refs.~\cite{Ejiri:2000fc,Irges:2009bi,deForcrand:2010be,Akerlund:2015poa,Knechtli:2016pph}. 
Yet another application comes from supersymmetric Yang-Mills theory and superstring theory;  
the compactification of the $d=10$ SU($N$) theory to $p$ dimensions leads to Yang-Mills theory with $d-p$ adjoint scalars, which is the bosonic part of the maximal Super-Yang-Mills theory describing $N$-coincident D$(p-1)$-branes. 
As demonstrated in Refs.~\cite{Aharony:2004ig,Aharony:2005ew}, such a theory is useful for understanding the phase diagram 
of $(p+1)$-dimensional Super-Yang-Mills, by interpreting the former to be the high-temperature limit of the latter. The main argument behind this connection is the fact that in this limit the fermions acquire a large thermal mass due to antiperiodic boundary conditions in the temporal direction, and therefore effectively do not contribute to some observables\footnote{Clearly, approximating SYM by just its bosonic content is limited to observables that are not sensitive to the effective number of degrees of freedom. In particular, the approximation fails for the Stefan Boltzmann limit of the free energy of SYM at high temperature. However, lattice studies in full QCD have indicated that the pure bosonic theory is able to offer quantitatively accurate descriptions for the Wilson loop expectation value at high temperatures \cite{Cheng:2008bs}.}

In all these cases, the $(\mathbb{Z}_N)^d$ center symmetry plays a crucial role. 
This symmetry is characterized by $W_i\to e^{2\pi i n_i/N}W_i$, $n_i\in\mathbb{Z}$, 
where $W_i$ is the Wilson line winding on the $i$-th dimension (see section \ref{sec:numerical_setup} for a lattice definition).
The center symmetry along each direction can be broken 
when that direction is small. Dimensional reduction can make sense after the center symmetry has been broken.

From string theory point of view, this transition can be interpreted 
as the black hole/black string transition or its higher-dimensional analogue \cite{Aharony:2004ig,Aharony:2005ew,Hanada:2007wn}.

In this paper we use the Hybrid Monte Carlo method for studying this theory.
Firstly, as a sanity check, we study the $d=10$ theory on an $N_t\times 1^9$ lattice, 
which reduces to a $(0+1)$-dimensional theory with 9 scalars in the continuum limit. 
We observe good agreement with previous simulation result  \cite{Aharony:2004ig,Kawahara:2007fn}. 

Next we consider the compactification to two dimensions, 
by taking the lattice size to be $N_t\times N_s\times 1^8$. 
The lattice spacing $a$ is taken to be the same for all directions.  
At sufficiently small lattice spacing, the theory reduces to two-dimensional Yang-Mills with 8 scalars. 
Although the scalars can acquire mass through radiative corrections, we expect this mass to grow only logarithmically with the lattice spacing
and hence it likely is not important for the small lattices used in this paper; 
we can expect our simulation results to be close to the massless theory, which is the bosonic part of maximally supersymmetric two-dimensional Yang-Mills theory. This begs the natural question: how close are the two theories, or in other words, how important are the effects from fermions? 
One clear difference is that the supersymmetric theory has only a deconfined phase, while the bosonic theory has both confined and deconfined phases. 
As we will see, deep in the deconfined phase, our simulation results for the bosonic theory are very close to the results for the supersymmetric theory obtained before by Catterall, Joseph and Wiseman \cite{Catterall:2010fx} (see also \cite{Giguere:2015cga} for other simulation results on the same theory and \cite{Joseph:2015xwa} for a review of this and related topics).

By using the same simulation code, it is possible to study classical real-time dynamics without fermions. The observation that bosonic and supersymmetric theories give close results suggests that the classical treatment could offer a potentially quantitatively good description in the deconfined phase.
  Previous studies in other dimensions, e.g. about the thermalization \cite{Asplund:2011qj,Asplund:2012tg,Aoki:2015uha} and scrambling \cite{Kunihiro:2010tg,Gur-Ari:2015rcq}, might be justified in this way. It would be interesting if various real-time aspects of string theory, such as the fast scrambling \cite{Sekino:2008he,Shenker:2013pqa,Maldacena:2015waa} and the black hole/black string topology change \cite{Gregory:1993vy,Choptuik:2003qd,Aharony:2004ig,Catterall:2010fx}, 
could be studied in a similar manner.

In the remainder of this work we study the dynamics of the theory on an $N_t\times N_s^3\times 1^6$ lattice, 
which can be thought to approximate ${\cal N}=4$ Super-Yang-Mills (SYM) in $(3+1)$ dimensions in the high-temperature limit. We map out the location of the deconfinement transition on various lattice sizes and calculate the corresponding thermodynamic pressure as a function of the lattice coupling. We then continue to demonstrate that simulating the full ten-dimensional theory is technically possible and then summarize and give our conclusions. Detailed treatment of two somewhat more technical aspects can be found in the appendices.

\section{Yang-Mills in Toroidal Compactification}
We consider the SU($N$) pure Yang-Mills theory on $d$-dimensional torus, whose action is defined by 
\begin{eqnarray}
\label{eq:action}
S=\frac{1}{4g_{(d)}^2}\sum_{i,j=1}^d\int d^dx {\rm Tr}F_{ij}^2, 
\end{eqnarray} 
where the field strength $F_{ij}$ is defined by $F_{ij}=\partial_i A_j-\partial_j A_i-i[A_i,A_j]$. 
Note that the coupling constant $g_{(d)}^2$ has a dimension of $({\rm mass})^{4-d}$. Also, note that this action uses holographic normalization convention, whereas standard QCD convention would have a prefactor of $\frac{1}{2 g_{(D)}^2}$ instead. 
Formally, the lattice regularization can be obtained by using the standard plaquette action, 
\begin{equation}
\label{eq:Laction}
S=\beta_{\rm eff} \sum_{x\in T^d,\Box} \left(1-\frac{1}{N}{\rm Re\ Tr\ } U_\Box\right)\,,
\end{equation}
where $\beta_{\rm eff}\equiv\frac{N a^{d-4}}{g_{(d)}^2}$ is the $d$-dimensional effective lattice coupling. Note that $\beta_{\rm eff}$ has mass dimension zero and that $\beta_{\rm eff}$ scales as $\propto N^2$ in the large $N$ limit.
At $d>4$, this theory is power-counting non-renormalizable, and it is most likely not well-defined in the continuum limit. 
Still, it can be treated as a cut-off theory, and there are efforts motivated by the extensions of the standard model with extra dimensions
(see e.g. in \cite{Ejiri:2000fc,Irges:2009bi,deForcrand:2010be,Akerlund:2015poa}). 

If we take the lattice to be T$^p\times 1^{d-p}$, Yang-Mills theory coupled to $d-p$ adjoint scalars can be obtained. Again, when $p>4$, the theory is not renormalizable. 
In this paper we concentrate on $p\le 4$ where the continuum theory is well-defined. 
Note that, unless counter-terms are added, the scalar masses receive radiative corrections so that the masses may become large in the continuum limit. On the lattice, this has been studied in the context of torelon masses (see for instance ~\cite{Michael:1986cj,DeGrand:1987bm,deForcrand:2010be}). One finds that heavy scalars effectively decouple from the theory, so that e.g. $d=5$ Yang-Mills reduces to usual four-dimensional Yang-Mills in the continuum limit except if the coupling and extra dimension size are fine-tuned \cite{deForcrand:2010be}.

In this work, we do not fine-tune the scalar masses but rather check if our results are sensitive to the number of scalars $d-p$ by performing lattice simulations with different values for the parent dimension $d$. If the scalars become very heavy, then one expects results to be insensitive to the choice of $d$ while conversely a sensitivity to $d$ implies that the scalars have not (completely) decoupled from the theory.

\subsection{Expected Phase Diagram and String Theory Interpretation}
In this section we review the previous numerical and analytical results concerning the phase diagrams. 
For simplicity we consider only the square tori, i.e. all compactification periods are taken to be the same value $L$. 

\begin{figure}[htbp]
  \centering
  \includegraphics[width=0.45\textwidth]{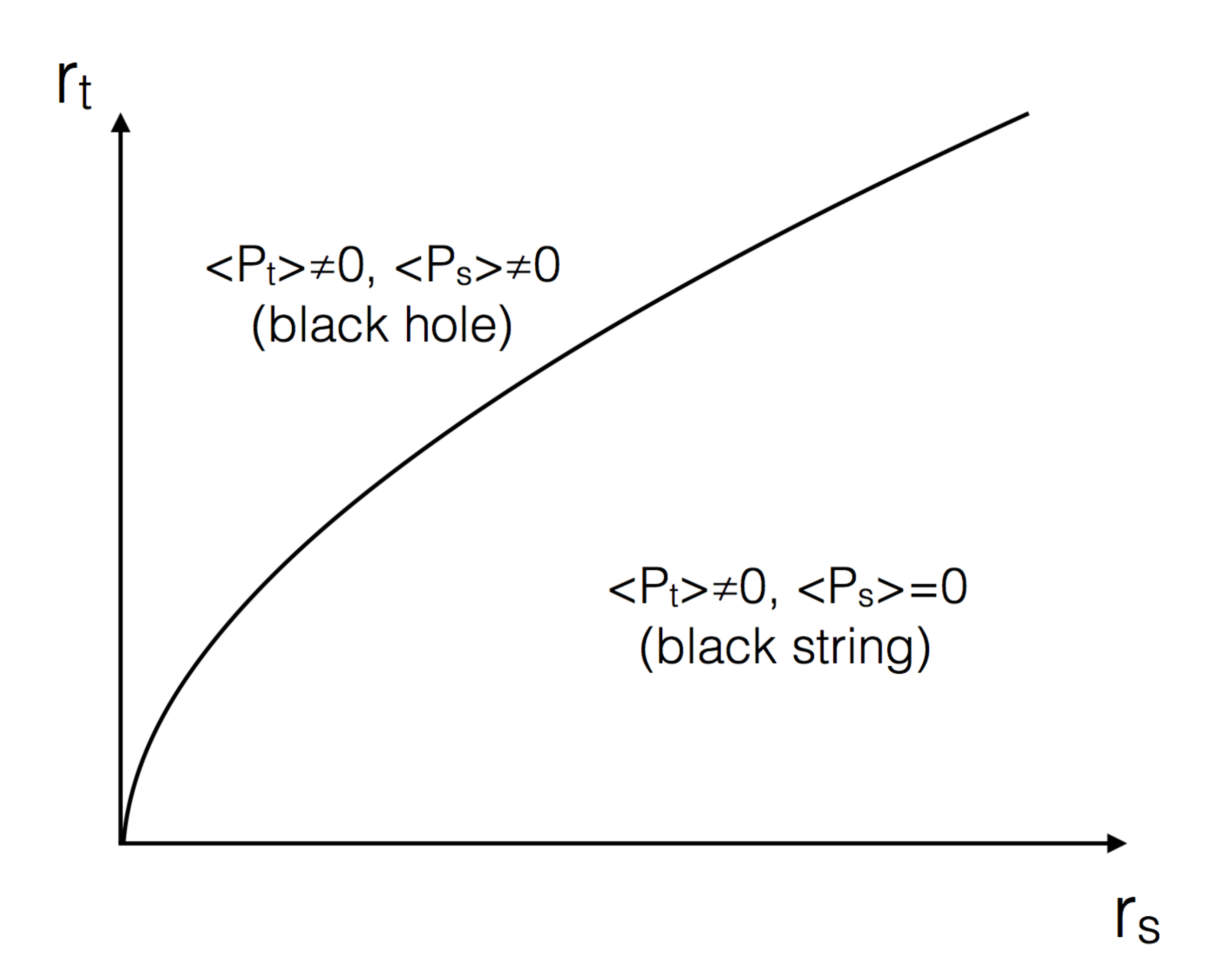}
  \hfill
  \includegraphics[width=0.45\textwidth]{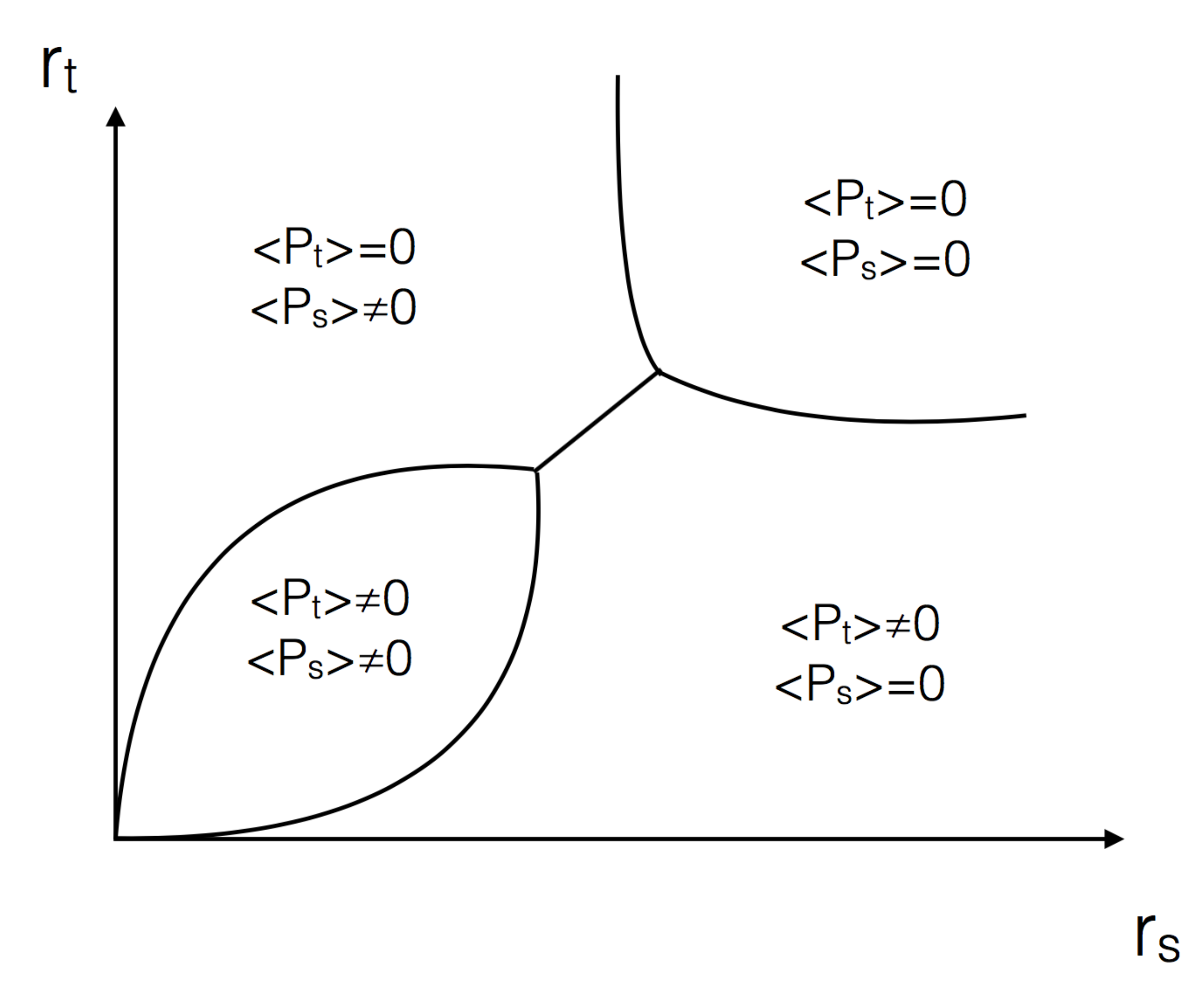}
  \caption{
  Conjectured phase diagram of 2d SYM (left) \cite{Aharony:2005ew} and 2d bosonic theory (right) \cite{Aharony:2005ew,Hanada:2007wn} at finite temperature, which is expected for massive scalars. 
  The present work provides the first numerical check on the conjectured bosonic phase diagram.  
 At the high-temperature region of SYM, two transitions (uniform to non-uniform, and non-uniform to localized) are expected \cite{Kawahara:2007fn,Mandal:2009vz}.
  }\label{Fig:2d-phase}
\end{figure}

Let us start with pure Yang-Mills without scalars. Narayanan and Neuberger \cite{Narayanan:2003fc,Narayanan:2005en}
have studied T$^3$ and T$^4$, taking the continuum and large-$N$ limits. 
(Note that, at finite $N$, the phase transition cannot take place at finite volume.)
It has been observed that the center symmetry breaks down sequentially as $({\mathbb Z}_N)^d\to ({\mathbb Z}_N)^{d-1}\to\cdots\to \{1\}$. One observable sensitive to this sequential breaking pattern is $P_i\equiv {\rm Tr}W_i/N$ since every time one of the $({\mathbb Z}_N)$'s breaks, one of $P_i$'s gets a nonzero expectation value. When the center symmetry is broken to $({\mathbb Z}_N)^{p}$, the system can be described approximately by Yang-Mills with $d-p$ scalars. 
Therefore, it is natural to expect the center breaking pattern \cite{Narayanan:2003fc,Narayanan:2005en} 
also for theories with adjoint scalars; see the right panel of Fig.~\ref{Fig:2d-phase} for $(1+1)$-d theory with scalars.

\begin{figure}[htbp]
  \centering
  \includegraphics[width=0.8\textwidth]{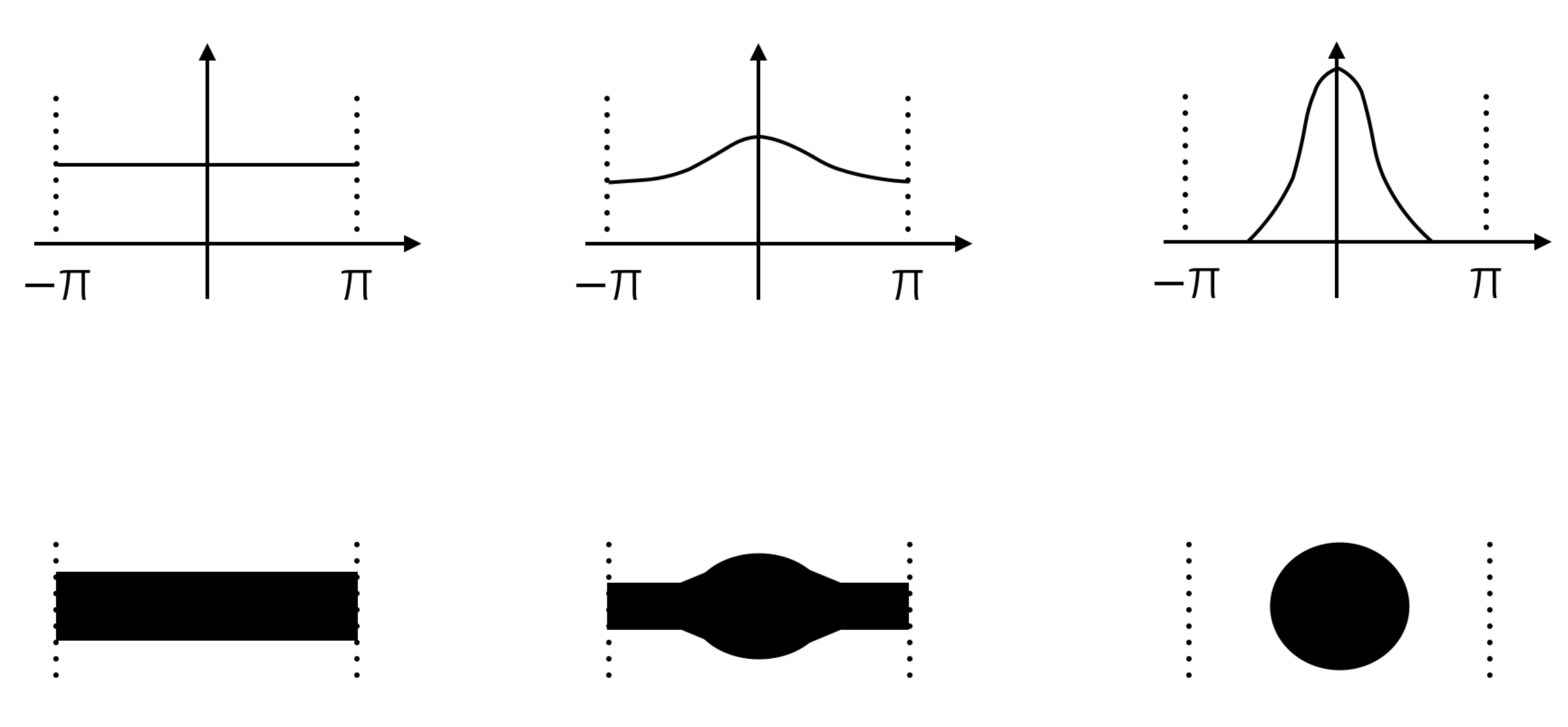}
  \caption{
  Possible Wilson line phase distributions in $(1+1)$-dimensional maximal SYM (top) and dual gravity interpretation (bottom).
  The Wilson line phases correspond to the positions of D0-branes in the T-dual picture on the gravity side.  
  The uniform distribution corresponds to the uniform black string (left), non-uniform distribution without gap at $\pm\pi$ corresponds to 
  the non-uniform string (middle), and the localized distribution corresponds to the black hole (right). 
  \label{Fig:BH-BS}}
\end{figure}

The case of $d=10$, $p\le 4$ theory can be regarded as the high temperature limit of maximally supersymmetric Yang-Mills theory in $(p+1)$ dimensions,   
which has an interpretation in terms of dual superstring theory. 
As a concrete example, let us first consider $d=10$, $p=1$ \cite{Aharony:2004ig,Aharony:2005ew}. 
The corresponding $(1+1)$-dimensional maximal SYM is dual to black 1-brane in type IIB superstring theory at low temperature and at large-$N$. 
The finite temperature and $1/N$ corrections correspond to stringy corrections. 
In the center-symmetric phase, the Wilson line phases are distributed uniformly (top-left of Fig.~\ref{Fig:BH-BS}), 
while in the center-broken phase a non-uniform distribution without gap at $\pm\pi$ (top-middle of Fig.~\ref{Fig:BH-BS})  
and the localized distribution (top-right of Fig.~\ref{Fig:BH-BS}) can exist.   
The Wilson line phases correspond to the positions of D0-branes in the T-dual picture on the gravity side; 
hence these three phases correspond to the uniform black string, the non-uniform black string\footnote{
Note that, in this context, `black string' does not mean the black 1-brane; it means the smeared, string-shaped distribution of D0-branes. 
} and the black hole (bottom of Fig.~\ref{Fig:BH-BS}). 
The thermodynamic study on the gravity side \cite{Aharony:2004ig,Aharony:2005ew} 
shows that there exist two phases at low temperature, 
the uniform black string phase at large volume and the black hole phase at small volume, which are separated by a first-order phase transition. 
The $(0+1)$-dimensional bosonic theory has a similar phase diagram at high temperature; the center symmetry is broken at small volume 
(high temperature, deconfined phase) and unbroken at large volume (low temperature, confined phase). 
However, unlike the low temperature region of the $(1+1)$-dimensional maximal SYM, 
the center-symmetric phase further splits into the non-uniform distribution without gap and the localized distribution \cite{Kawahara:2007fn,Mandal:2009vz}. 
By combining these arguments, the phase diagram of the $(1+1)$-dimensional maximal SYM becomes should qualitatively look like
the left panel of Fig.~\ref{Fig:2d-phase}.

The same analysis applies to $(2+1)$- and $(3+1)$-dimensional maximal SYM theories, just by considering 
the higher-dimensional counterparts of the black string. The thermodynamic analysis on the gravity side \cite{Hanada:2007wn} gives
the center breaking pattern observed in the bosonic theory, namely 
$({\mathbb Z}_N)^2\to {\mathbb Z}_N\to \{1\}$ and $({\mathbb Z}_N)^3\to ({\mathbb Z}_N)^2\to {\mathbb Z}_N\to \{1\}$. 
Therefore it is natural to expect $p$-dimensional Yang-Mills with adjoint scalars, which resembles the high temperature region of $(p+1)$-dimensional 
SYM, has the center breaking pattern $({\mathbb Z}_N)^p\to ({\mathbb Z}_N)^{p-1}\to\cdots\to \{1\}$. 
In \cite{Aharony:2005ew} a part of the phase diagram of the 2d Yang-Mills with heavy adjoint scalars has been studied analytically, 
and the same conclusion have been obtained.

\section{Numerical setup}\label{sec:numerical_setup}

Let us first discuss the setup for solving the {\it classical} Yang-Mills equations of motions in ``real'' time for a system in $(d+1)$ dimensional Minkowski space-time. It will turn out that this formalism can be re-interpreted as the molecular dynamics evolution of a Hybrid Monte Carlo simulation of the {\it quantum} Yang-Mills system for a system in $d$ dimensional Euclidean space-time. This re-interpretation will be described in more detail below. For now, let us consider the case of classical Yang-Mills.

Let us consider an SU($N$) gauge field $A_\mu(x)$ in $(d+1)$ dimensional Minkowski space-time which can be expanded in terms of the generators $T^a$ of the corresponding SU($N$) Lie algebra \cite{Montvay:270707},
\begin{equation}
A_\mu(x)= -i g \sum_a A_\mu^a(x) T^a\,,
\end{equation}
where the $T^a$ are traceless hermitian complex $N\times N$ matrices and $g$ is the bare gauge coupling and the coefficient functions $A_\mu^a(x)$ are real. We take them as being normalized as ${\rm Tr}\left(T_a T_b\right)=\frac{\delta_{ab}}{2}$. (See appendix \ref{app1} on how an algorithm for constructing the generators for arbitrary $N$). Here and in the following Greek indices $\mu=(0,i)$ denote $(d+1)$ space-time with Minkowski signature whereas Latin indices from the middle of the alphabet such as $i,j,k,\ldots$ denote $d$ dimensional Euclidean space and Latin indices from the beginning of the alphabet such as $a,b,c,\ldots$ denote $N^2-1$ dimensional group space. For $(d+1)$ dimensional Minkowski space-time, the mostly plus metric convention will be used ($g_{\mu\nu}={\rm diag}(-,+,+,+,\ldots)$).
The equations of motion of this gauge field are described by the Lagrangian density ${\cal L}=-\frac{1}{4}F_{\mu\nu}^a F^{\mu \nu a}$ where
\begin{equation}
\label{eq:fmunu}
F_{\mu\nu}^a=\partial_\mu A_\nu^a-\partial_\nu A_\mu^a+g f^{abc} A_\mu^b A_\nu^c\,,
\end{equation}
is the associated field strength tensor and $f^{abc}$ are the real and completely antisymmetric structure constant of the SU($N$) gauge group that obey the relation $\left[T_a,T_b\right]=\sum_c i f^{abc} T_c$ \cite{Montvay:270707}. The equations of motion for the gauge field can be written as $D_\mu^{ab}F^{\mu\nu a}=0$ where
\begin{equation}
D_{\mu}^{ab}\equiv \partial_\mu \delta^{ab}-i f^{abc} A_\mu^c\,.
\end{equation}

Using the temporal gauge $A_0^a=0$, the Lagrangian density gives rise to the classical Hamiltonian density ${\cal H}$ via the usual Euler-Lagrange equations,
\begin{equation}
\label{hden}
{\cal H}=\frac{1}{4}E_i^a E_i^a+\frac{1}{8}F_{ij}^a F_{ij}^a\,,
\end{equation}
where $E_i^a\equiv \partial_0 A_i^a=F_{0i}^a$ can be interpreted as the electric fields. Note that the prefactors of $\frac{1}{4},\frac{1}{8}$ are unconventional from a QCD perspective, but chosen in order to make contact with holographic literature. In addition to the Hamiltonian equations of motion for $A_i$, one has to ensure the Gauss-law constraint 
\begin{equation}
D_i^{ab}E_i^a={\rm const}\,.
\end{equation}
\subsection{Real Time Formulation on the Lattice}
In order to solve the above real time Hamiltonian equations while preserving gauge invariance it is useful to make use of standard lattice formulations of gauge fields. To this end, one replaces continuum Euclidean space by an isotropic cubic lattice such that $x^i=a \hat{x}^i$ with $\hat{x}$ taking on integer values and $a$ being the (spatial) lattice spacing. The continuum gauge field variables $A_i(x)$ are replaced by link variables \hbox{$U_i(x)=e^{a A_i(x)}=e^{-ig a A_i^a(x) T^a}$} which are elements of the SU($N$) Lie group and live on links between lattice sites $\hat{x}^i$. More specifically, by $U_i(x)$ we mean the link pointing from site $\hat{x}^j$ to site $\hat{x}^j+\hat{e}_{i}$ and obeying $U_i^\dagger(x)=e^{i g a A_i^a(x) T_a}=e^{-a A_i(x)}$.
With this identification, we may use the standard single plaquette definition for the magnetic contribution to the Hamiltonian density (\ref{hden})
\begin{equation}
\label{eq:magcontr}
\frac{1}{8}F_{ij}^a F_{ij}^a\rightarrow \frac{N}{g^2 a^4} \sum_\Box \left(1-\frac{1}{N}{\rm Re\ Tr\ } U_\Box\right)\,,
\end{equation}
where $U_\Box$ is a plaquette variable defined through a spatial loop on the lattice \cite{Montvay:270707,Bodeker:1999gx}:
\begin{equation}
U_{\Box,ij}=U_i(x)U_j(x+i)U_i^\dagger(x+j)U_j^\dagger(x)\,,
\end{equation}
and $\sum_\Box$ in Eq.~(\ref{eq:magcontr}) denotes the sum over all spatial loops on the lattice starting from site $\hat{x}^i$ with only one orientation, e.g. \cite{Montvay:270707}
\begin{equation}
\sum_\Box\equiv \sum_{1\leq i < j \leq d}\,.
\end{equation}
The Hamiltonian density for the system is thus given as 
\begin{equation}
\label{hdendisc}
{\cal H}=\frac{1}{4}E_i^a E_i^a+\frac{N }{g^2 a^4}\sum_\Box\left(1-\frac{1}{N}{\rm Re\ Tr} U_{\Box,ij}\right)\,.
\end{equation}
We can calculate the lattice equations of motion from this Hamiltonian. First note that the Hamiltonian equation $\frac{d A_i^a(x)}{dt}=\frac{\partial {\cal H}}{\partial E_i^a(x)}=E_i^a$. We can use this to calculate the time derivative of the link variable $U_i$ as 
\begin{equation}
\label{eq:Utimedep}
\frac{d U_i(x)}{d t}=-i g a T^a E_i^a U_i(x)=\frac{i E_i(x)}{a} U_i(x)\,,
\end{equation}
when using the definition $E_i(x)=-g a^2 T^a E_i^a(x)$. Discretizing time in units of $t=a \Delta t\, \hat{t}$ with $\hat{t}$ an integer, the update rule for the link variables becomes (cf.~\cite{Bodeker:1999gx})
\begin{equation}
\label{eq:Uevol}
U_i(\hat{x},\hat{t}+1)=e^{i \Delta t E_i(x)}U_i(\hat{x},\hat{t})\,.
\end{equation}
Note that while analytic expressions for the exponential of a matrix exist notably for $N=2$, this is not the case for arbitrary $N$. For this reason, we approximate the exponential via its power series, keeping a finite number of terms. Approximating the exponential will in general lead to violations of unitarity; we monitor the total unitarity violation incurred in the simulation and have adjusted the number of terms used in the power series expansion of the exponential in order to ensure the violations to be on the order of double precision machine accuracy (in practice, we found that keeping the first 8 terms in the series expansion is sufficient).

To obtain the update rule for the electric field, note that the total energy $H(t)$ of the system should be conserved as a function of time. On a lattice, the total energy is given by
\begin{equation}
H(t)=\int d^dx {\cal H}=a^d \sum_x {\cal H}\,,
\end{equation}
where $\sum_x$ denotes the sum over all lattice sites.

Requiring $\dot{H}\equiv \frac{d H(t)}{dt}=0$ one finds 
\begin{equation}
0=\sum_{x}\sum_{i=1}^d\left[E_i^a(x)\dot{E}_i^a(x)-\frac{1}{g^2 a^4}\sum_{j=1}^{d}{\rm Re\ Tr\ }\frac{d U_{\Box,ij}}{dt}\right]\,,
\end{equation}
where $\sum_{i,j>i}=\frac{1}{2}\sum_{i,j}$ was used. $U_{\Box,ij}$ consists of the product of four link variables $U_i$, for each of which the time derivative can be calculated by using (\ref{eq:Utimedep}). Using then the cyclic property of the trace, the symmetry under the interchange $i\leftrightarrow j$, as well as the possibility of shifting the site index $x+i\rightarrow x$ because of the overall sum over all lattice sites one finds that $\dot{H}=0$ implies
\begin{equation}
\label{eq:masterE}
\dot{E}_i^a(x)=\frac{2}{g a^3}\sum_{|j|\neq i}{\rm Im\ Tr}\left[T^a U_i(x)S_{ij}^\dagger(x)\right]\,,
\end{equation}
(cf. \cite{Gottlieb:1987mq,Bodeker:1999gx,Berges:2008zt}), where the index $j$ runs over both positive and negative values and the gauge staple $S_{ij}$ is defined as \cite{Bodeker:1999gx}
\begin{equation}
S_{ij}(x)=U_j(x) U_i(x+j) U_j^\dagger(x+i)\,.
\end{equation}
Note that for negative values of $j$, a gauge link is traversed in the opposite direction, e.g.
\begin{equation}
U_{-j}(x)=U_j^\dagger(x-j)\,.
\end{equation}

It is advantageous for numerical stability to record the link variables on integer time steps which storing the electric field on half integer time-steps (leap-frog algorithm). Eq.~(\ref{eq:masterE}) may be recast into an evolution equation
\begin{equation}
\label{eq:Eevol}
E_i(\hat{x},\hat{t}+\frac{1}{2})=E_i(\hat{x},\hat{t}-\frac{1}{2})-\Delta t\sum_{|j|\neq i} {\rm Adj}\left[U_i(\hat{x},\hat{t})S_{ij}^\dagger(\hat{x},\hat{t})\right]\,,
\end{equation}
where ${\rm Adj}\left[M\right]\equiv - \frac{i}{2}\left[M-M^\dagger-\frac{1}{N}{\rm Tr}\left(M-M^\dagger\right)\right]$ for SU($N$).

 Real time lattice evolution is done by evolving initial conditions for the field $U_i,E_i$ using the set of equations (\ref{eq:Uevol},\ref{eq:Eevol}). The total energy of the system can be written as
\begin{equation}
\label{eq:Etot}
H(\hat{t})=\frac{N a^{d-4}}{g^2}\sum_x\left[\frac{{\rm Tr}\left[\left(E_i(\hat{x},\hat{t}+\frac{1}{2})+E_i(\hat{x},\hat{t}-\frac{1}{2})\right)^2\right]}{8 N}+\sum_\Box \left(1-\frac{1}{N}{\rm Re\ Tr} U_{\Box,ij}(\hat{t})\right)\right]\,,
\end{equation}
where the square in the first term really denotes $E_i^2\equiv E_i E_i^\dagger$. This Hamiltonian is conserved up to order ${\cal O}(a^2)$ violations as a function of time. The Gauss law constraint on the lattice can be monitored by calculating 
\begin{equation}
\label{eq:gausslaw}
G(\hat{t})=\sum_i \left(E_i(\hat{x},\hat{t}+\frac{1}{2})-U_i^\dagger(\hat{x}-\hat{e}_i,\hat{t}) E_i(\hat{x}-\hat{e}_i,\hat{t}+\frac{1}{2})U_i(\hat{x}-\hat{e}_i,\hat{t})\right)\,,
\end{equation}
and making sure that $G(\hat{t})\simeq 0$ up to machine precision errors.
\subsection{Hybrid Monte Carlo Simulations}
An efficient algorithm to perform importance sampling for a probability density $e^{-S}$ with $S$ the classical action of the theory is provided by Hybrid Monte Carlo (HMC) simulations \cite{Montvay:270707}. In the HMC simulations, the standard heat-bath algorithm is combined with classical evolution to give rise to new gauge field configurations. It is straightforward to implement an HMC algorithm based on the above classical evolution equations. Consider again the classical action in $d$ dimensional Euclidean space-time (\ref{eq:action}) implemented on the lattice (\ref{eq:Laction}).
Define fictitious ``conjugate'' momenta $E_i^a$ for the classical evolution of the gauge field configurations in Langevin time $t$. The Hamiltonian density for this formulation will be given by Eq.~(\ref{hdendisc}), and the equations of motion are given by equations (\ref{eq:Utimedep}) and (\ref{eq:masterE}), respectively. The particular HMC algorithm we are using in the following is then defined as
\begin{enumerate}
\item
Choose initial link configurations to be trivial, $U_i(x)={\bf 1}$
\item
Throw momenta $\tilde E_i^a$ randomly from a Gaussian ensemble
\item
Scale momenta as $\tilde E_i^a\rightarrow E_i^a=\tilde E_i^a \sqrt{\frac{N}{\beta_{\rm eff,d}}}$ such that the Hamiltonian and Equations of motion are again given by (\ref{eq:Etot}),(\ref{eq:Uevol}),(\ref{eq:Eevol})
\item
Calculate the total ``energy'' of the system $H_0$ given by Eq.~(\ref{eq:Etot})
\item 
Evolve momenta $E_i$ by half a step
\item
Evolve $U_i$ and $E_i$ using (\ref{eq:Uevol}),(\ref{eq:Eevol}) with step size $\Delta t$ for a certain number of steps $n_{\rm steps}$. For the last step, the momenta are only evolved half a step.
\item
Calculate the total ``energy'' of the system $H_f$ by evaluating Eq.~(\ref{eq:Etot})
\item
The new configuration $U_i^\prime$ is accepted with probability
\begin{equation}
P={\rm min}\left\{1,e^{-H_f+H_0}\right\}
\end{equation}
If the new configuration is rejected, load the previous gauge field configuration
\item
Repeat from (2)
\end{enumerate}

A standard choice for the time step $\Delta t$ is to adopt a value such that acceptance of new configurations is around 80 percent. This allows efficient generation of new configurations while avoiding auto-correlations. In our simulations, we have found that even for acceptance rates that approach unity, autocorrelations for the observables we are interested in are not a problem. Thus we only decrease $\Delta t$ if acceptance rates become too low. The number of steps $n_{\rm step}$ has to be chosen sufficiently large such that individual configurations thermalize. In practice, we have found that $n_{\rm step}\times \Delta t\gtrsim 10$ is required to ensure full thermalization of individual trajectories.

\subsection{Toroidal compactification}

 Since the action (\ref{eq:action}) must be dimensionless this leads to the mass dimensions of the gauge field and $d$ dimensional coupling
\begin{equation}
\left[A_i^a\right]=\frac{d-2}{2}\,,\quad \left[g_{(d)}\right]=\frac{4-d}{2}\,.
\end{equation}
In particular, this implies that 
\begin{equation}
\left[A_i\right]=\left[ g_{(d)} A_i^a\right]=1\,,
\end{equation}
so the combination $a A_\mu$ appearing in the gauge links $U_i$ is dimensionless as it should be. Reducing the theory to $p$ Euclidean space-time dimensions by taking the number of sites to be one in $d-p$ of the $d$ directions leads to a lattice action (\ref{eq:Laction}) with effective coupling 
\begin{equation}
  \beta_{{\rm eff},p}\equiv \frac{N a^{d-4}}{g_{(d)}^2}=\frac{N a^{p-4}}{g_{(p)}^2}=\frac{N^2 a^{p-4}}{\lambda_{(p)}}\end{equation}
and the sum over x as well as the sum over plaquettes now are restricted to the non-compactified direction. In the limit $a\rightarrow 0$ this essentially turns the gauge fields along the compactified direction into scalars. Note that $\beta_{\rm eff}$ can be expressed in terms of the $p$ dimensional 't Hooft coupling $\lambda\equiv N g^2$, which we will use in the following.

We conclude by remarking that the simulation code thus described is publicly available as a service to the community \cite{codedown}.


\subsection{Observables}
\label{sec:lattice}
As observables we will consider Wilson/Polyakov loops which are defined as $W_i={\cal P}e^{i \oint dx A_i(x)}$ for a loop in the ${\rm i}^{\rm th}$ (Euclidean) direction, where ${\cal P}$ denotes path ordering. On the lattice, this is easily implemented through a product of link matrices
\begin{equation}
W_i=\prod_{m=1}^{m=n_i} U_i(m)\,,
\end{equation}
where $n_i$ denotes the number of lattice sites in the direction i. We will distinguish the case where $i=t$ (the Euclidean ``time'' direction) and $i=(x,y,z)=s$ (one of the Euclidean ``space'' directions), corresponding to the case of Polyakov and Wilson loops, respectively. Of particular interest will be the expectation value of the absolute value of the trace (cf. Ref.~\cite{Catterall:2010fx}),
\begin{equation}
  \label{eq:polyakov}
  \langle P_i \rangle = \left\langle \left|\frac{1}{N}{\rm Tr}\, W_i\right|\right\rangle\,,
\end{equation}
averaged over directions transverse to $i$ and over all configurations. Also, we will study the distribution of eigenvalues $\{e^{i \theta_1},e^{i \theta_2},\ldots,e^{i \theta_N}\}$ of the Wilson/Polyakov loops on the unit circle. The phases of the eigenvalues are required to sum to $2\pi$ times the integer winding number for the SU($N$) gauge group. We will ignore non-trivial topologies in this study and re-center the phases of the eigenvalues so they sum to zero if a non-trivial winding number occurs.
The distribution of eigenvalue phases on the interval $[-\pi,\pi]$ will then be of interest as an order parameter for the systems considered here.


Another observable will be the equation of state, e.g. the functional dependence of the energy density $\varepsilon$ on the temperature T. Given a partition function $Z=\int dU e^{-S}$, the energy density and pressure $P$ are calculated as derivatives w.r.t. temperature and volume $V$,
\begin{eqnarray}
  \label{eq:therm1}
  \varepsilon&=&-\frac{1}{V}\frac{\partial \ln Z}{\partial (1/T)}=\frac{T}{V}\left\langle a_t \frac{\partial S}{\partial a_t} \right\rangle \nonumber\\
  P&=&T\frac{\partial \ln Z}{\partial V}=\frac{-T}{(d-1)V}\left\langle a_s \frac{\partial S}{\partial a_s}\right\rangle\,.
\end{eqnarray}
Note that derivatives w.r.t temperature $T=\frac{1}{N_t a_t}$ and volume $V=(N_s a_s)^{d-1}$ for a lattice with a fixed number of sites $N_t,N_s$ in the temporal and spatial directions have been recast as derivatives w.r.t. $a_t,a_s$, the (in principle different) lattice spacings in these directions. For an anisotropic lattice with different lattice spacings in the temporal and spatial directions the action is given by
\begin{equation}
  S=\frac{N^2 a_s^{d-3}}{\lambda_{(d)} a_t}\sum_{x} R_{\rm temp}
  +\frac{N^2 a_t a_s^{d-5}}{\lambda_{(d)}}\sum_{x} R_{\rm spat}\,,
  \end{equation}
where the ``temporal'' and ``spatial'' plaquettes are given as (cf. \cite{Montvay:270707}).
\begin{equation}
  R_{\rm temp}\equiv \sum_{i=1}^{d-1}\left(1-\frac{1}{N}{\rm Re\ Tr\ } U_{ti}\right)\,,\quad
  R_{\rm spat}\equiv \sum_{i=1}^{d-1}\sum_{j>i}^{d-1}\left(1-\frac{1}{N}{\rm Re\ Tr\ } U_{ij}\right)\,.
\end{equation}
Toroidal compactification simply amounts to the replacement $\frac{a_s^{d}}{\lambda_{(d)}}\rightarrow \frac{a_s^{p}}{\lambda_{(p)}}$. 
Away from $p=4$, the coupling is dimensionful and can be scaled away in all physically relevant observables. Thus only the explicit dependence on the lattice spacings contributes to the derivatives in (\ref{eq:therm1}). Setting $a_t=a_s=a$ in the end (isotropic lattice) leads to 
\begin{eqnarray}
  \varepsilon&=&\frac{T}{V} \frac{N^2 a^{p-4}}{\lambda_{(p)}}\sum_x\left(-\langle R_{\rm temp}\rangle +\langle R_{\rm spat}\rangle\right)\,,\\
  P&=&\frac{T}{(p-1)V} \frac{N^2 a^{p-4}}{\lambda_{(p)}}\sum_x\left(-\langle R_{\rm temp}\rangle (p-3)-\langle R_{\rm spat}\rangle (p-5)\right)\,.\nonumber
\end{eqnarray}
As a special case, let us mention that for $p=1$ (matrix model quantum mechanics), we have
\begin{equation}
  \label{eq:E1d}
  \epsilon V = E =\frac{N^2 a^3 T}{\lambda_{(1)}} \sum_x\left(-\langle R_{\rm temp}\rangle +\langle R_{\rm spat}\rangle \right)\,.
  \end{equation}
For $p=4$, in particular, it is advantageous to consider a different strategy to calculate thermodynamic properties. Namely, note that a different definition of the pressure is $P=\frac{T}{V}\ln Z$. Considering
\begin{equation}
  \frac{\partial \ln Z}{\partial \beta_{\rm eff,p}}=-\left\langle \frac{\partial S}{\partial \beta_{\rm eff,p}}\right\rangle
  =-\sum_{x}\left(\langle R_{\rm temp}\rangle +\left\langle R_{\rm spat}\right\rangle \right)\,.
\end{equation}
Integration of this result leads to the pressure. However, care must be taken to remove the zero-temperature contribution. In practice, this is done by subtracting $\langle R_{\rm temp}\rangle +\langle R_{\rm spat}\rangle$ calculated on a lattice with $N_s^D$ sites at each value of $\beta_{\rm eff}$ \cite{Montvay:270707}, 
\begin{eqnarray}
  \frac{P}{T^{D-p}}&=&\int_{\beta_0}^\beta d\bar \beta \Delta S(\bar\beta)\,,\nonumber\\
  \Delta S(\bar \beta)&=&N_t^{D-p}\left[S(N_s,N_s,\bar \beta)-S(N_t,N_s,\bar \beta)\right]\,,\label{eq:deltaS}\\
  S(N_t,N_s,\bar\beta)&\equiv& \frac{1}{N_s^{D-1}N_t} \sum_x \left.\left(\langle R_{\rm temp}\rangle +\langle R_{\rm spat}\rangle\right)\right|_{\bar \beta}\,.\nonumber
  \end{eqnarray}
\newpage


\section{Results}
\subsection{$d=10$ $\rightarrow$ $1d$: Matrix Model Quantum Mechanics}
\label{sec:1d}

\begin{figure}[t]
  \centering
  \includegraphics[width=0.49\textwidth]{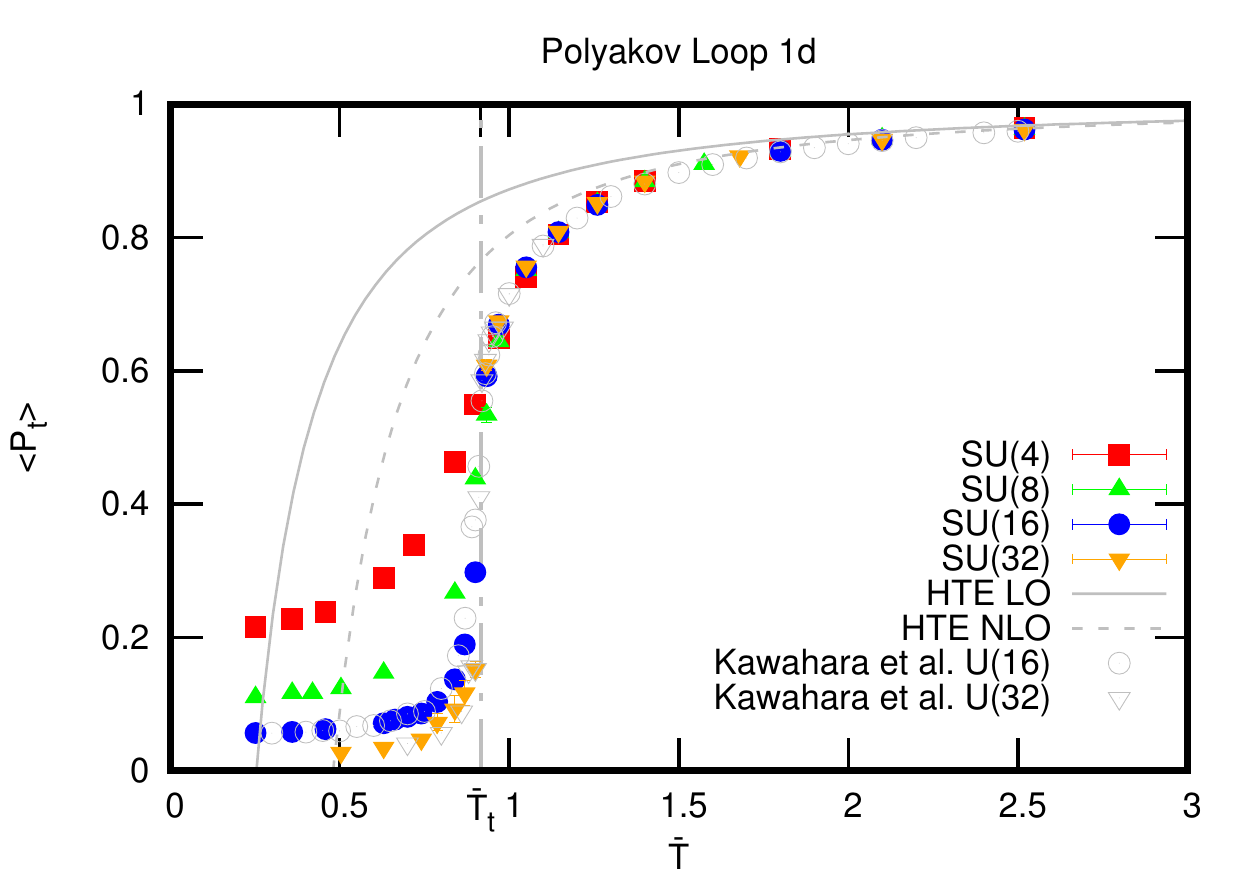}
  \hfill
  \includegraphics[width=0.49\textwidth]{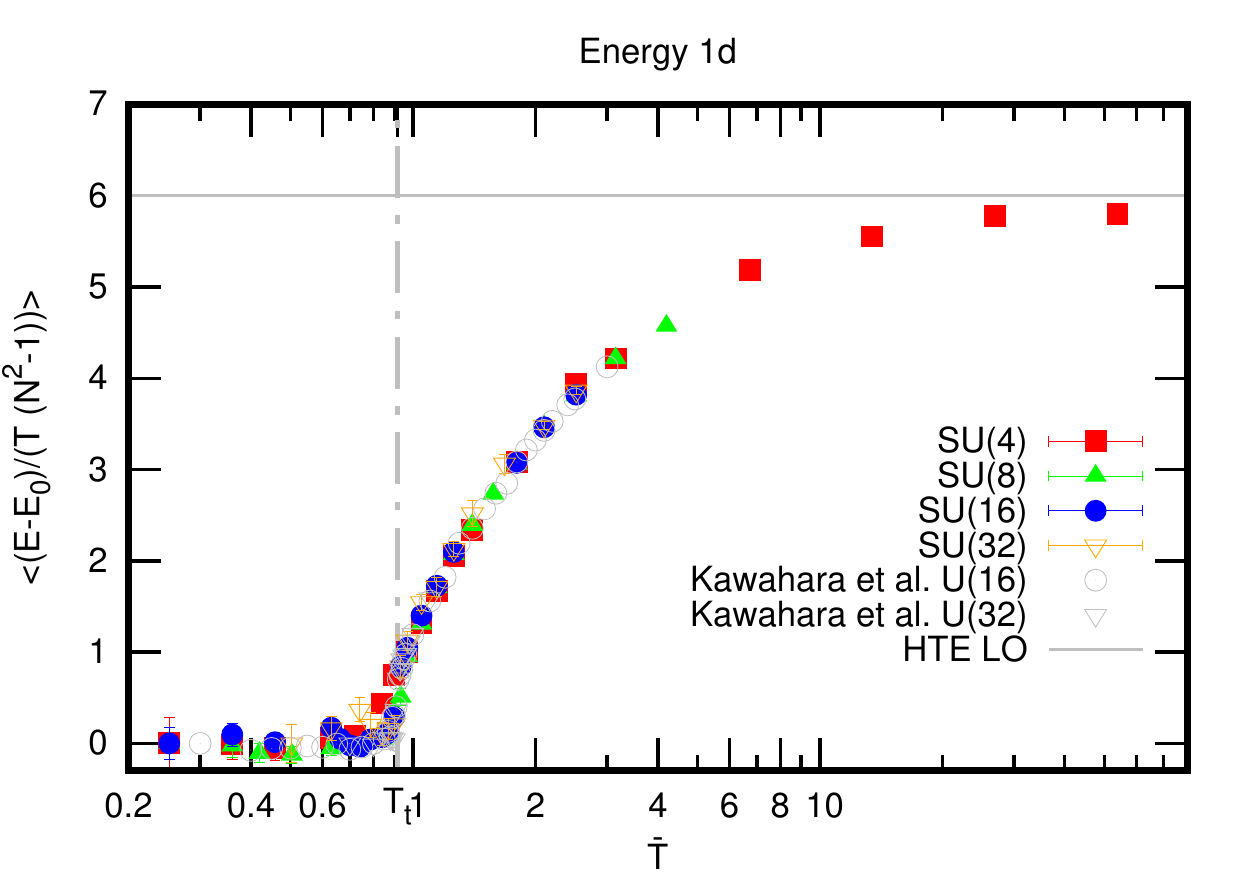}
  \caption{\label{fig:one}
    Results from simulating 10d SU($N$) compactified to 1d (matrix model quantum mechanics) for $\beta_{\rm eff}=\frac{N^2}{0.04^3}$ (effective lattice spacing $\bar a=0.04$) for various values of $N$, and $\beta_{\rm eff}=\frac{N^2}{0.01^3}$ for $N=4$. Shown are the expectation value of the Polyakov loop $<P_t>$ and the expectation value of the energy $E$ (minus the energy at zero temperature $E_0$ which is calculated from the lattice with the largest value of $N_t$) as a function of the dimensionless temperature $\bar T$ (full symbols). For comparison, analytic results from a high-temperature-expansion (HTE) \cite{Kawahara:2007ib}, as well as previous lattice simulation results are shown \cite{Kawahara:2007fn}. The dash-dotted line indicates the location of $T_t$ and is a guide to the eye.   }
\end{figure}

Starting with ten-dimensional SU($N$) Yang-Mills and compactifying down to $p=1$ dimension leads to a Lagrangian corresponding to bosonic matrix model quantum mechanics. This theory has been well-studied both analytically as well as numerically \cite{Aharony:2004ig,Kawahara:2007fn,Mandal:2009vz}. We conduct simulations for fixed values of $\beta_{\rm eff,1}=\frac{N^2}{\lambda_{(1)}a^3}$ on an isotropic lattice, where the continuum limit $a\rightarrow 0$ corresponds to $\beta_{\rm eff,1}\rightarrow \infty$.  Using the dimensionful 't Hooft coupling $\lambda_{(1)}$, we can consider dimensionless quantities such as $a \lambda^{1/3}_{(1)}$, $T \lambda^{-1/3}_{(1)}$, which we will denote by a bar, e.g. $\bar{a}\equiv a \lambda^{1/3}_{(1)},\bar{T}\equiv T \lambda^{-1/3}_{(1)}$.
%
%
%
%
The results for the Polyakov loop $\langle P_t\rangle$ and the energy $E$, obtained by evaluating (\ref{eq:polyakov},\ref{eq:E1d}) on the lattice configurations, are shown in Fig.~\ref{fig:one}. Also shown in Fig.~\ref{fig:one} are numerical results from Ref.~\cite{Kawahara:2007fn}, as well as high-temperature-expansion (HTE) analytic results from Ref.~\cite{Kawahara:2007ib}.
As can be seen from Fig.~\ref{fig:one}, there is very good agreement between the present study and previous results. The difference between simulating U($N$) and SU($N$) gauge theories does not seem to be relevant. 

\begin{figure}[t]
  \includegraphics[width=0.49\textwidth]{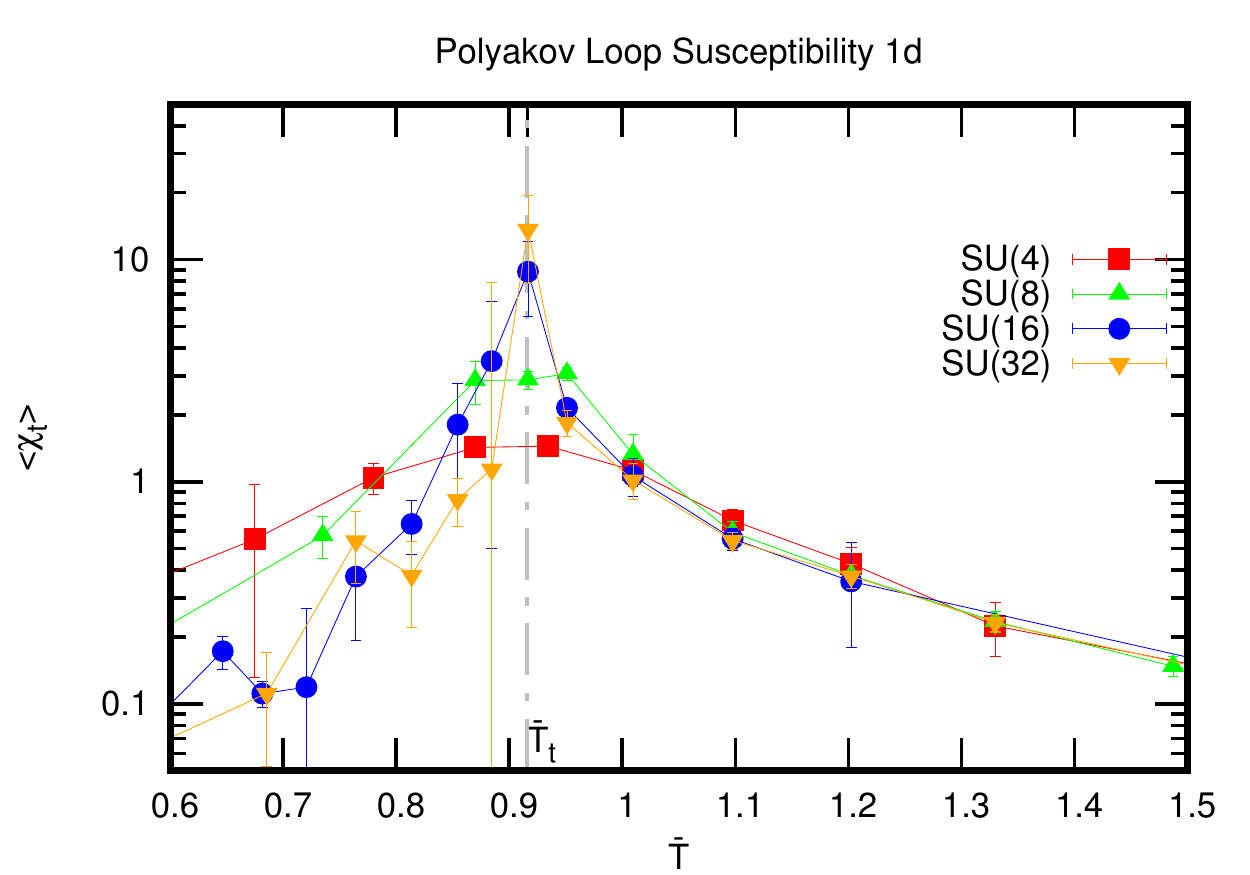}
  \hfill
  \includegraphics[width=0.49\textwidth]{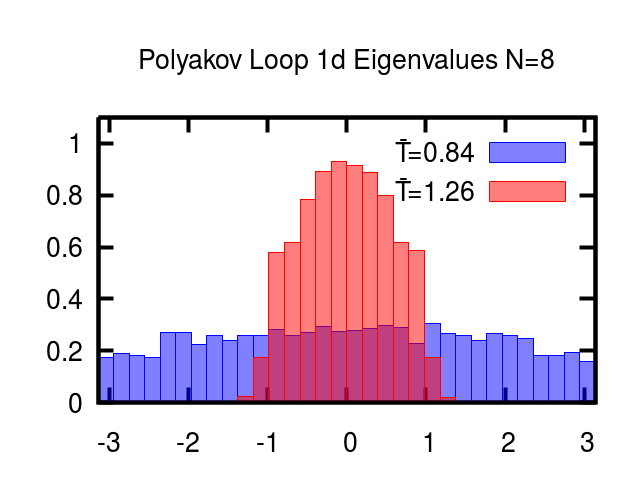}
  \caption{\label{fig:EV1d} Left: 1d Polyakov loop susceptibility as a function of temperature for various $N$ and $\beta{\rm eff}=\frac{N^2}{0.04^3}$. For large N, and this resolution, we find the susceptibility peak to lie at $\bar T_t=0.916(30)$. Right: The eigenvalue distribution for the 1d Polyakov loop is found to change qualitatively from non-uniform ($\bar T<\bar T_t$) to localized ($\bar T>\bar T_t$).}
\end{figure}

We observe a rapid change in the expectation value of the Polyakov loop as a function of temperature. Defining the susceptibility $\chi_{t}\equiv \frac{d \langle P_{t,s}\rangle}{d T}$, we define $\bar{T}_t$ by the location of the peak in $\chi_t$ (see Fig.~\ref{fig:EV1d}). For a resolution of $\beta_{\rm eff}=\frac{N^2}{0.04^3}$ we find $\bar{T}_t\simeq 0.916(30)$; increasing the resolution to $\beta_{\rm eff}=\frac{N^2}{0.02^3}$ we find $\bar{T}_t\simeq 0.885(15)$, consistent with previous results \cite{Aharony:2004ig, Kawahara:2007fn}. As can be seen from Fig.~\ref{fig:EV1d}, the peak in the susceptibility and hence the change in the Polyakov loop expectation value becomes more pronounced as the number of colors is increased, and also corresponds to a change in the eigenvalue distribution of the Polyakov loop (see Fig.~\ref{fig:EV1d}). As the temperature is decreased, the distribution of the Polyakov loop eigenvalues becomes more and more uniform, but because of the limited statistics for this study we cannot rule out the possibility of a second phase transition from non-uniform to a fully uniform distribution.

Comparing the value of $\bar{T}_t$ with the results for the energy in Fig.~\ref{fig:one}, we are led to identify $\bar{T}_t$ with the critical temperature for the confinement-deconfinement transition in the bosonic 1d Matrix Model. Since results shown in Fig.~\ref{fig:EV1d} indicate that the Polyakov loop eigenvalue distribution also changes qualitatively from localized ($\bar{T}>\bar{T}_t$) to non-uniform ($\bar{T}<\bar{T}_t$) at the same critical temperature $\bar{T}_t$, we are led to identify the confined phase with the non-uniform eigenvalue phase and the deconfined phase with the localized eigenvalue phase.


\subsection{Arbitrary $d$ $\rightarrow$ 1d and 2d: Sensitivity to Number of Scalars}

\begin{figure}[t]  
  \includegraphics[width=0.49\textwidth]{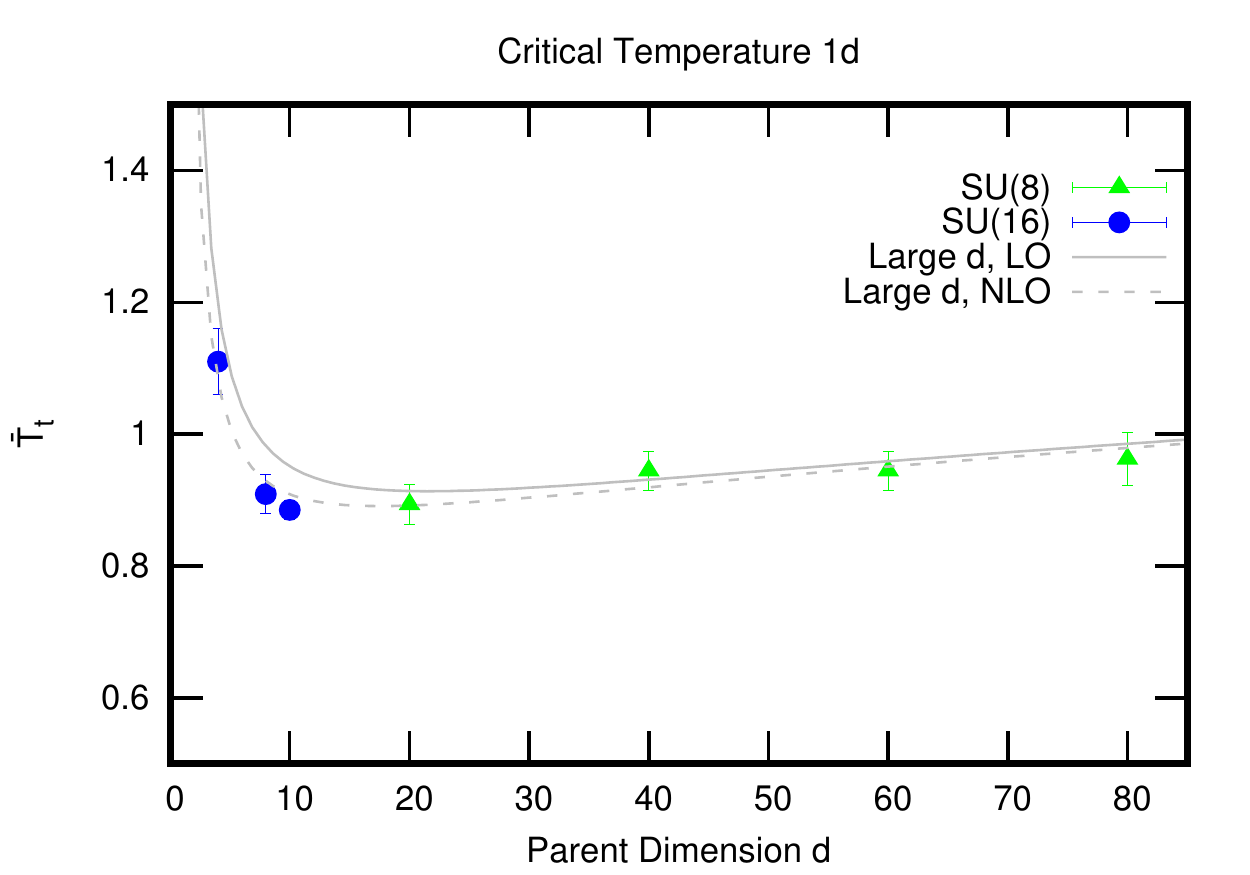}
  \hfill
   \includegraphics[width=0.49\textwidth]{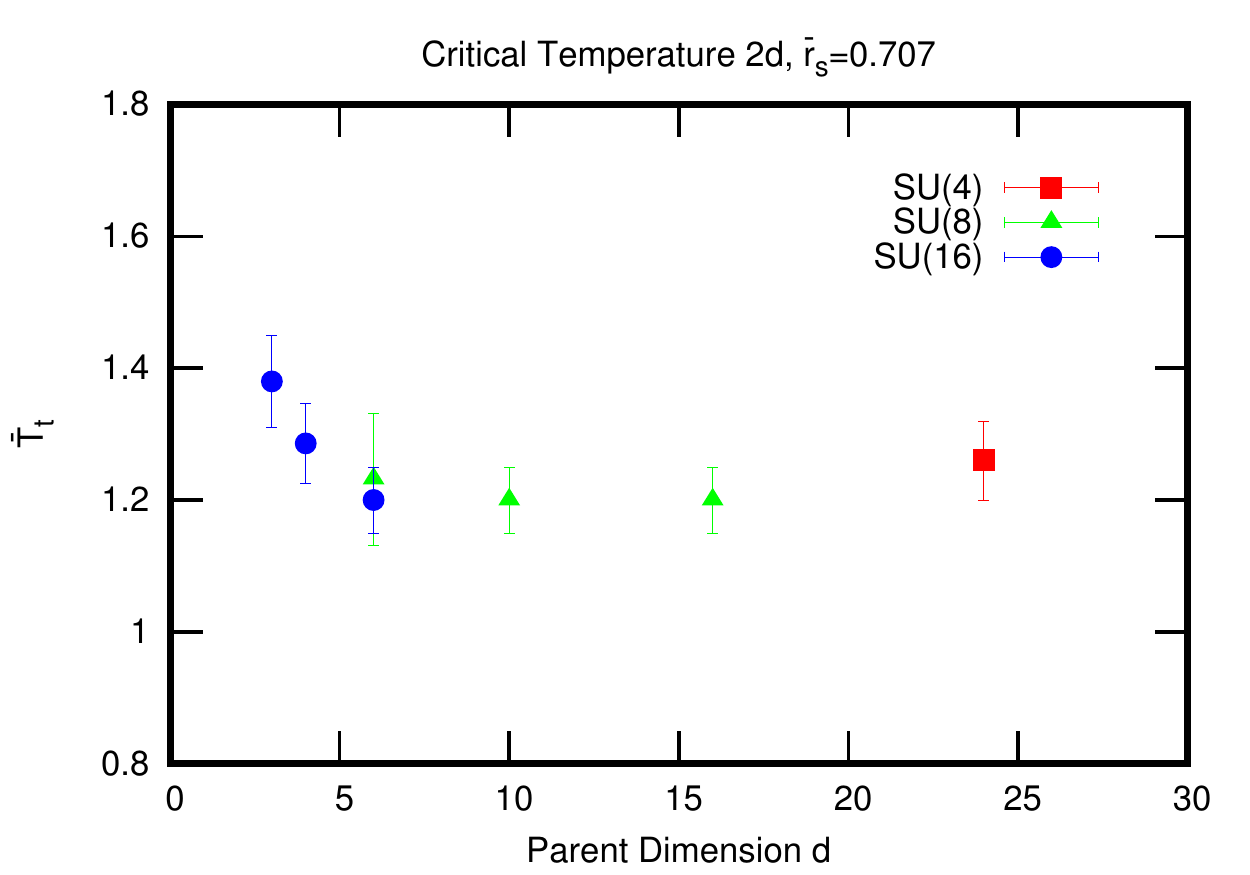}
  \caption{\label{fig:1dDscan} Dependence on the critical temperature for the Polyakov loop (defined via the peak in the susceptibility) as a function of varying parent dimension $d$. Left: Reduction to 1d. For comparison, the analytic results obtained in a large $d$ approximation from Ref.~\cite{Mandal:2009vz} are shown. Right: Reduction to 2d with $\bar{r}_s=0.707$. The behavior of $\bar{T}_t$ is qualitatively similar to 1d.}
\end{figure}

To check for the sensitivity to the number of scalars, we perform simulations with varying parent dimension $d$, always compactifying down to $p=1$. Similar studies have been performed by other groups before \cite{Janik:2000tq,Azuma:2014cfa}. We monitor the Polyakov loop expectation value and locate the peak in its susceptibility, defining the location of the critical (deconfinement) temperature $T_t$. We find indications that as $d$ is increased above $d=10$, the susceptibility peak and hence the strength of the transition at fixed $N$ becomes more pronounced. Conversely, for $d<10$, the transition seems to weaken at fixed $N$ and we have to increase $N$ to clearly identify a peak in the Polyakov loop susceptibility. The results for $\bar{T}_t$ versus parent dimension $d$ are shown in Fig.~\ref{fig:1dDscan}. For comparison, also the analytic results for $\bar{T}_t$ obtained in a large $d$ expansion from Ref.~\cite{Mandal:2009vz} are shown. We find remarkably good agreement with the next-to-leading order analytic result even for $d=4$. This agreement (and the implied dependence of $\bar{T}_t$ on the effective number of scalars $d-1$) suggests that in our simulations the scalars have not decoupled from the theory.

Similar findings hold true for the case of the reduction from $d=10$ to $p=2$ (2d). The results for $\bar{T}_t$ exhibit a clear dependence on the parent dimension $d$, which is qualitatively similar to the trend seen in the 1d case. We are not aware of any analytic calculations in the large $d$ limit for this case.

\subsection{$d=10$ $\rightarrow$ 2d: Yang-Mills on a circle}

Compactification of the 10d theory onto $(1+1)$ dimensions corresponds to Yang-Mills theory at finite temperature compactified on a circle.  Using the two-dimensional 't Hooft coupling $\lambda_{(2)}=\frac{N^2}{\beta_{\rm eff,2} a^2}$ we follow Ref.~\cite{Catterall:2010fx} to define a dimensionless circle radius $\bar{r}_s=N_s a \lambda_{(2)}^{1/2}$ and a ``temporal'' radius $\bar{r}_t=N_t a \lambda_{(2)}^{1/2}=1/\bar{T}$. In terms of these radii, the supersymmetric (SYM) version of this system is conjectured to have three possible phases, corresponding to situations which the Wilson loop eigenvalue distribution is localized, non-uniform and uniform. Holography has been used to study this system in the strong coupling (low temperature) limit, predicting the existence of a Gregory-Laflamme transition occurring at $\bar{r}_s^2\simeq 2.29 \bar{r}_t$, cf. Refs.~\cite{Aharony:2004ig,Aharony:2005ew,Catterall:2010fx} which is characterizing the change from the uniform to localized phase (the non-uniform phase is expected to be thermodynamically disfavored, cf.~\cite{Aharony:2004ig,Catterall:2010fx}).  At weak coupling (high temperatures), a deconfinement-confinement transition occurring at $\bar{r}_s^3\simeq 1.35 \bar{r}_t$ has been found in previous studies \cite{Aharony:2004ig,Kawahara:2007ib,Mandal:2009vz}. The high temperature result can be gleaned from results in section \ref{sec:1d} through identifying $T\rightarrow \lambda_{(2)}^{1/2}/\bar{r}_s$ and $\lambda_{(1)}=\lambda_{(2)}/a\rightarrow =\lambda_{(2)}^{3/2}/\bar{r}_t$ since the temporal circle becomes very short and temporal and spatial elements trade their respective meaning. In this manner it is obvious that our results in the high temperature limit must match the previous finding of $\bar{r}_s^3=\frac{\bar{r}_t}{(0.885(15))^3}\simeq 1.44(7) \bar{r}_t$.

The phase diagram of the 2d theory differs between the supersymmetric theory and the purely bosonic theory (see Fig.~\ref{Fig:2d-phase}). The supersymmetric theory should always be in the deconfined phase characterized by 
a nonzero Polyakov loop expectation value. The bosonic theory studied in this work is rather different: first of all, the phase diagram must be invariant under the exchange of the temporal and spatial circles, 
$\bar{r}_t\leftrightarrow \bar{r}_s$. By regarding this theory as the high-temperature limit of 3d SYM, 
the dual gravity calculation valid at the low-temperature region suggests a form of the phase diagram 
shown in the right panel of Fig.~\ref{Fig:2d-phase} \cite{Hanada:2007wn}. 
The conjectured phase diagram for the purely bosonic theory has not been confirmed previously by numerical simulations, except for the case of toroidal compactification of the toroidally compactified $d=4$ Yang-Mills theory for $\bar{r}_t=\bar{r}_s=\bar{r}_y=\bar{r}_z$ where a symmetry breaking pattern of
$(\mathbb{Z}_N)^4\to(\mathbb{Z}_N)^3\to(\mathbb{Z}_N)^2\to\mathbb{Z}_N\to\{1\}$ has been observed \cite{Narayanan:2005en}.



\begin{figure}[t]
  \centering
  \includegraphics[width=0.49\textwidth]{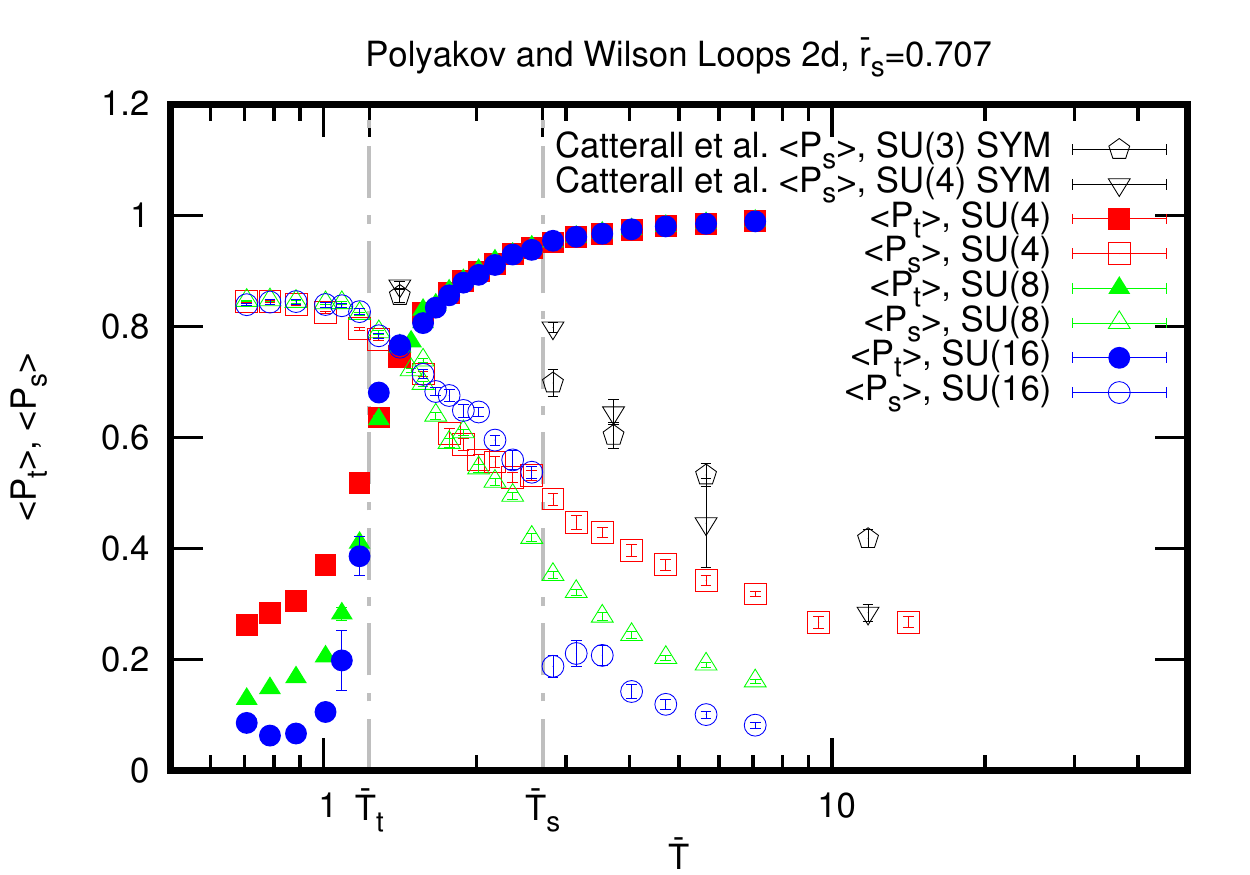}
  \hfill
  \includegraphics[width=0.49\textwidth]{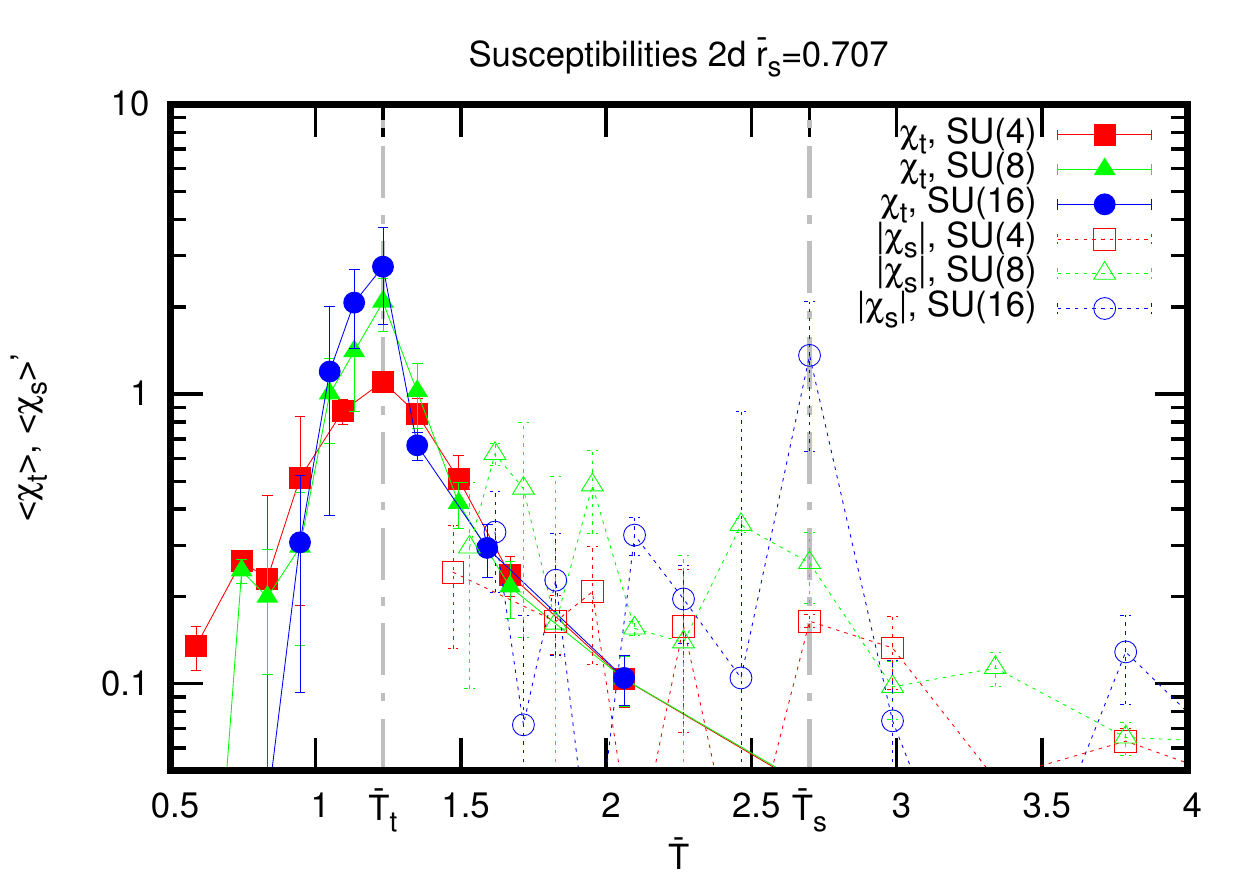}
  \caption{\label{fig:two} Left: Polyakov and Wilson loop for pure SU($N$) as a function of temperature for $\bar{r}_s=0.707$ (combined data with $\beta_{\rm eff}=\frac{N^2}{0.0707^2}$ for $\bar{T}<1.6$ and $\beta_{\rm eff}=\frac{N^2}{0.0353^2}$, respectively) compared to results for SYM for N=3,4 from Ref.~\cite{Catterall:2010fx}. Right: Polyakov and Wilson loop susceptibilities in pure SU($N$) as a function of temperature for for fixed circle radius $\bar{r}_s=0.707$. The dash-dotted lines indicate the location of $\bar{T}_{t,s}$ where the corresponding susceptibilities peak and are a guide to the eye.}
\end{figure}

We first study the Wilson and Polyakov loop expectation values for fixed spatial circle radius as a function of temperature in Fig.~\ref{fig:two}. We find that both the Wilson and Polyakov loops exhibits a rapid change as a function of temperature as $N$ is increased. Defining again the susceptibilities $\chi_{t,s}\equiv \frac{d \langle P_{t,s}\rangle}{d T}$, at any given spatial radius $\bar{r}_s$ for large enough $N$ we observe a peak at specific (different) temperatures $\bar{T}_{t,s}(\bar{r}_s)$ (see figure \ref{fig:two}). The peak for $\chi_t$ is visible for any $N\geq 4$ and becomes more pronounced as $N$ is increased, while the peak for $\chi_s$ only emerges for $N\geq 16$. Using the peak position from $\chi_t$ for $N\geq 8$ to define $\bar{T}_t$ and the peak for $\chi_s$ for $N=16$ to defined $\bar{T}_s$ we find $\bar{T}_t(\bar{r}_s\simeq 0.7)\simeq 1.232(10)$ and $\bar{T}_s(\bar{r}_s\simeq 0.7)\simeq 2.7(3)$. 

Also shown in Fig.~\ref{fig:two} are the results for SYM theory (gauge theory plus fermions) from Ref.~\cite{Catterall:2010fx}. At high temperatures the anti-periodic boundary conditions for the Fermions in SYM implies that they acquire a large mass and effectively decouple from the theory. Thus one expects good agreement between the full SYM theory and the pure bosonic SU($N$) theory simulations for high temperatures, while deviations are expected at low temperatures\footnote{The SYM simulations from Ref.~\cite{Catterall:2010fx} are based on a different discretization scheme, and require stabilization of flat directions in the scalar potential. Without supersymmetry, these flat directions are lifted at the quantum level, which is why stabilization is not needed in simulations of the purely bosonic theory. The fact that quantum effects lift the flat directions can be seen by calculating the interaction between eigenvalues. Essentially the same calculation has been performed in Ref.~\cite{Bhanot:1982cm} for a different motivation.}. Fig.~\ref{fig:two} seems to corroborate this expectation. We have performed a more extensive comparison between the SYM and pure SU($N$) gauge theory for a same-size lattice in appendix \ref{app2}, from which we expect quantitative agreement between the SYM and pure gauge theory for $T\geq 5 \lambda_{(2)}^{1/2}$.

\begin{figure}[t]
  \includegraphics[width=0.49\textwidth]{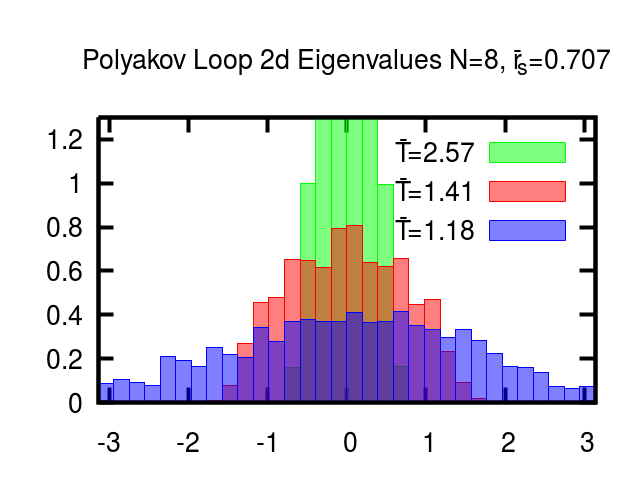}
  \hfill
  \includegraphics[width=0.49\textwidth]{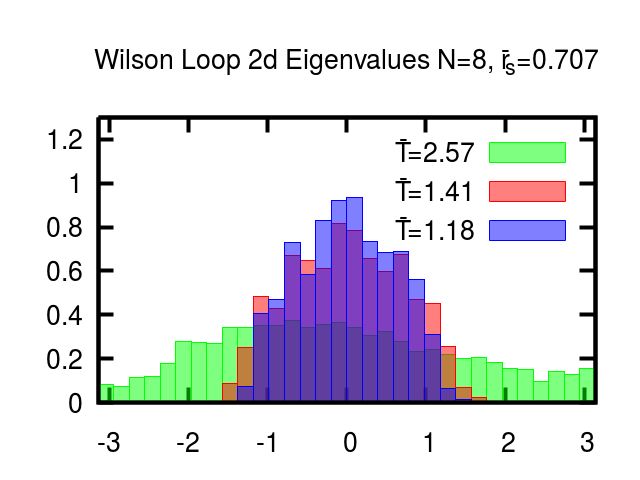}
  \caption{\label{fig:EV2d}  The eigenvalue distribution for the Polyakov loop (left) and Wilson loop (right). Both are for the 2d theory for $N=8$ and a circle radius of $\bar{r}_s=0.707$, calculated at $\beta_{\rm eff}\simeq \frac{N^2}{0.07^2}$ and $\beta_{\rm eff}\simeq \frac{N^2}{0.035^2}$, respectively. Note the change in the Polyakov loop behavior above/below $\bar{T}_t\simeq 1.232(10)$ and the Wilson loop behavior above/below $\bar{T}_s\simeq 2.7(3)$.}
\end{figure}

The different temperatures $\bar{T}_{t,s}$ are accompanied by corresponding changes in the distribution of eigenvalues of the Polyakov and Wilson loops (see Fig.~\ref{fig:EV2d}). Starting from low temperatures $\bar{T}<\bar{T}_t$ we find the distribution of the Polyakov loop eigenvalues to be non-uniform, but non-gapped, and the Wilson loop eigenvalues to be localized. For $\bar{T}_t<\bar{T}<\bar{T}_s$ the Polyakov loop and Wilson loop eigenvalues are both localized, and for $\bar{T}>\bar{T}_s$ the Polyakov loop eigenvalues are localized and the Wilson loop eigenvalues are non-uniform.

\begin{figure}[t]
  \centering
  \includegraphics[width=0.49\textwidth]{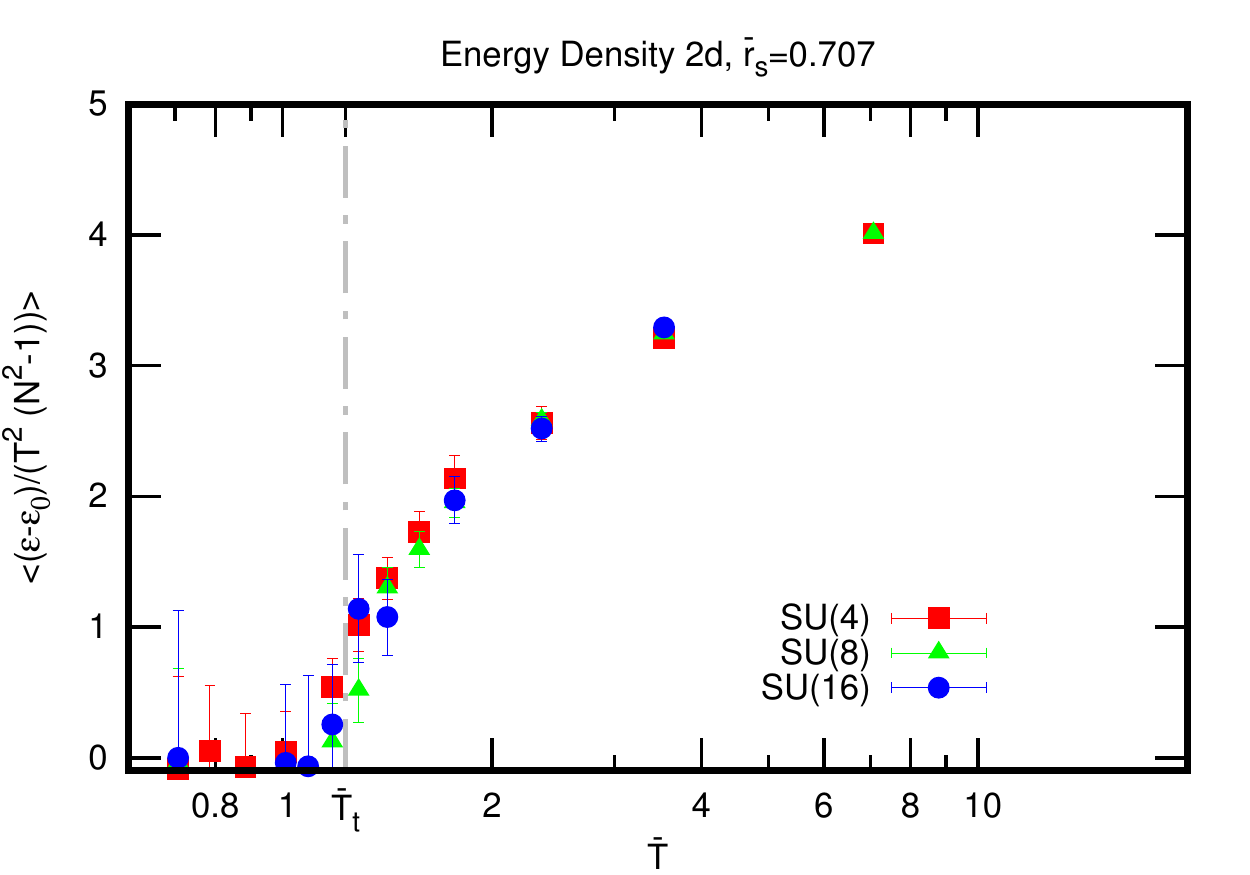}
\hfill
 \includegraphics[width=0.49\textwidth]{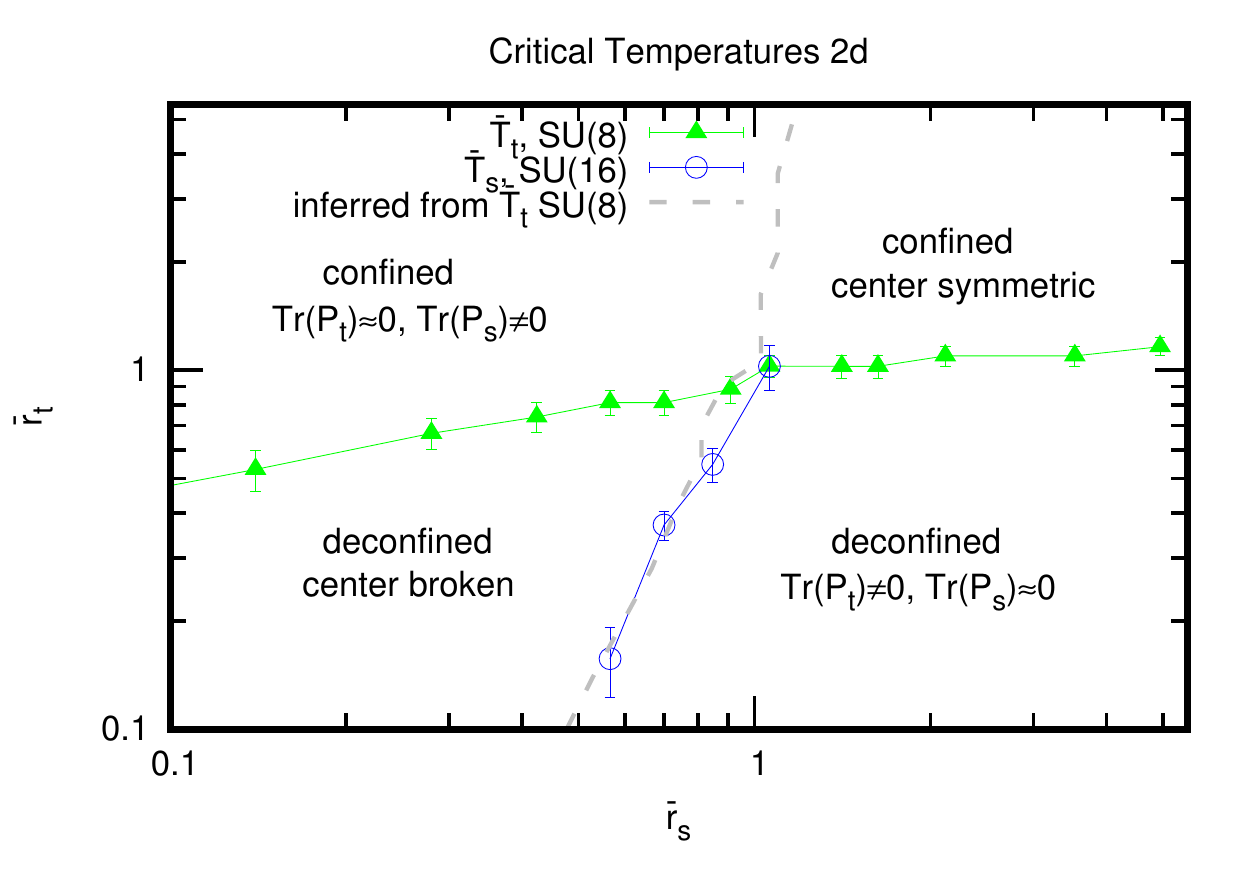}
 \caption{\label{fig:three} Left: 2d Energy density for $\bar{r}_s=0.707$ for various values of $N$, calculated for $\beta_{\rm eff}=\frac{N^2}{0.07^2}$. The vertical line labeled $\bar{T}_t$ indicates the peak position of the Polyakov loop and is a guide to the eye. Right: critical temperatures $\bar{T}_t,\bar{T}_s$ for the Polyakov loop and Wilson loop for N=8 and N=16, respectively, as a function of spatial circle radius $\bar{r}_s$.  The line labeled 'inferred' are results for $\bar{T}_t$ from SU(8) and switching $\bar{r}_s\leftrightarrow \bar{r}_t$.}
  \end{figure}

Comparing the temperatures $\bar{T}_{t}$ with the energy density as a function of temperature shown in Fig.~\ref{fig:three}, we are led to identify this temperature with the deconfinement-confinement transition as was the case for the 1d compactification. As can be seen from Fig~\ref{fig:two}, the deconfinement-confinement transition temperature $\bar{T}_t$ is clearly separated from the temperature $\bar{T}_{s}$. Because of the mapping of the high-temperature behavior and the behavior of the one-dimensional theory discussed in section \ref{sec:1d}, we identify $\bar{T}_{s}$ with a critical temperature signaling a Gregory-Laflamme (GL) type instability. This feature of having distinct confinement and GL critical temperatures is unique to the pure gauge SU($N$) theory considered here and is not expected to occur for the full SYM theory. In 3d and 4d cases, the confined phases can exist when the theories are compactified to 
two- and three-spheres \cite{Witten:1998qj}.

Also shown in Fig.~\ref{fig:three} are the locations of the critical temperatures for the Polyakov loop ($\bar{T}_t$) and Wilson loop ($\bar{T}_s$) as a function of the length of the spatial circle $\bar{r}_s$. The different transition temperatures correspond to boundaries of the phases with two unbroken center symmetries, one unbroken and one broken center symmetry, respectively, and both center symmetries broken. For large values of the spatial circle radius $\bar{r}_s \gg 1$ we find that the value of $\bar{T}_t(\bar{r}_s)$ becomes approximately independent of $\bar{r}_s$, consistent with the expectation from the infinite volume limit. However, note that $\bar{r}_s\gg 5$ would be necessary to unambiguously demonstrate this volume independence. 
We find qualitative agreement with the conjectured bosonic phase diagram in Fig.~\ref{Fig:2d-phase}. Close to $\bar{r}_t,\bar{r}_s\simeq 1$, Fig.~\ref{Fig:2d-phase} suggests the presence of a single transition line (``neck'') connecting the deconfined, center symmetry broken phase to the deconfined, center symmetry restored phase. On this line, one expects the center symmetry to to be partially broken, and we address this question in the following subsection.

\subsection{Sequential Breaking of $\mathbb{Z}_N$ symmetries in 2d and 4d}

\begin{figure}[t]
  \centering
  \includegraphics[width=0.49\textwidth]{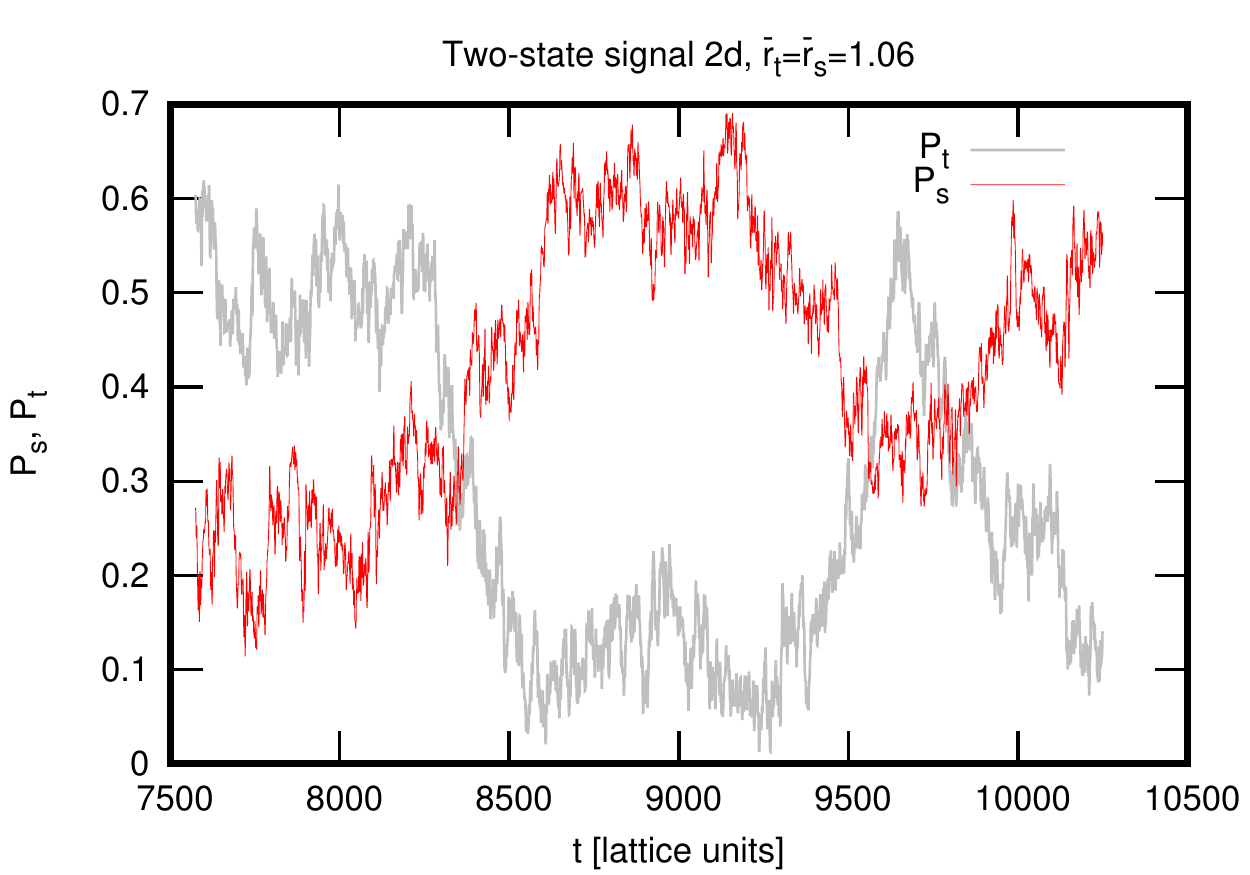}

 \caption{\label{fig:2s} Time history of $P_t,P_s$ for $N=16$ and $r_t=r_s$. Note that the time length of a single Monte Carlo trajectory is 25 in lattice units. }
  \end{figure}

\begin{figure}[t]
  \centering
  \includegraphics[width=0.49\textwidth]{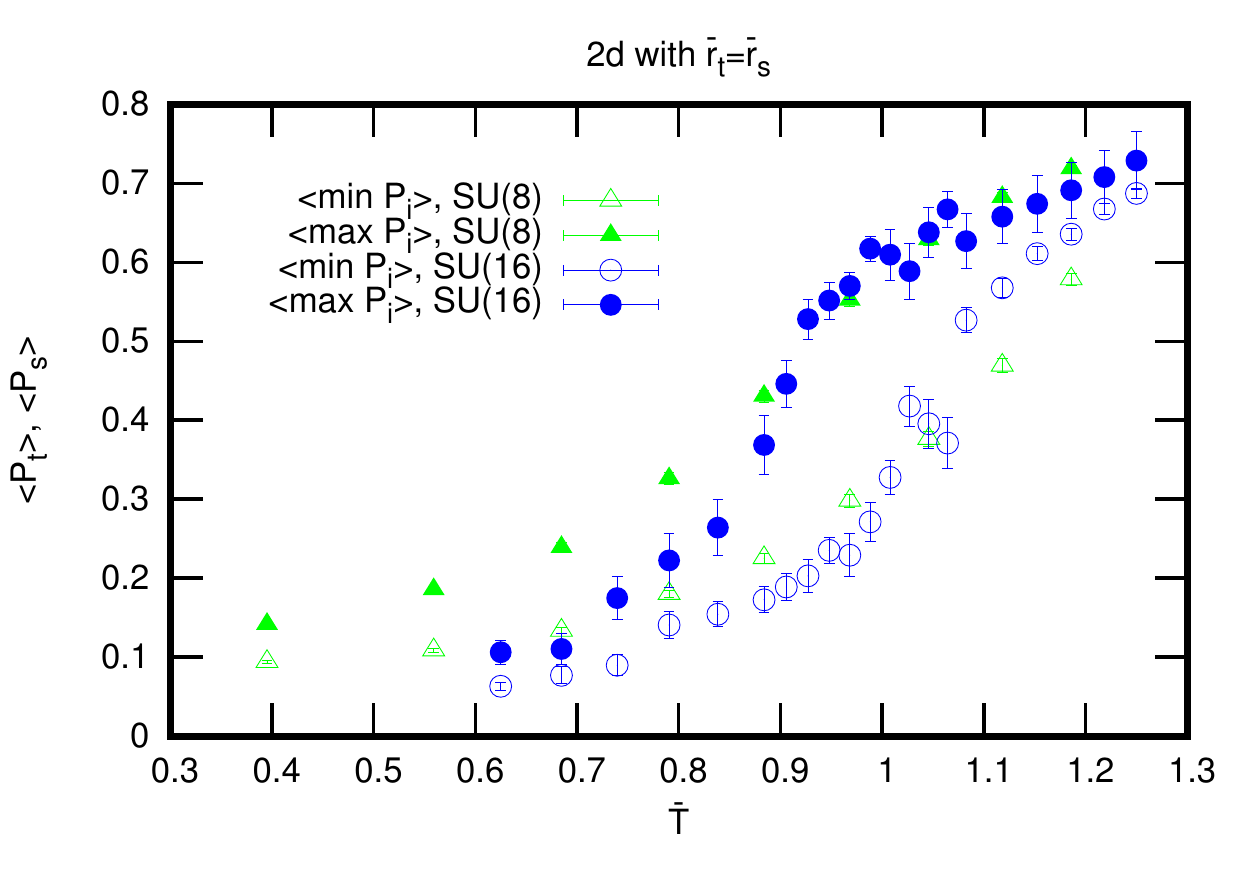}
\hfill
 \includegraphics[width=0.49\textwidth]{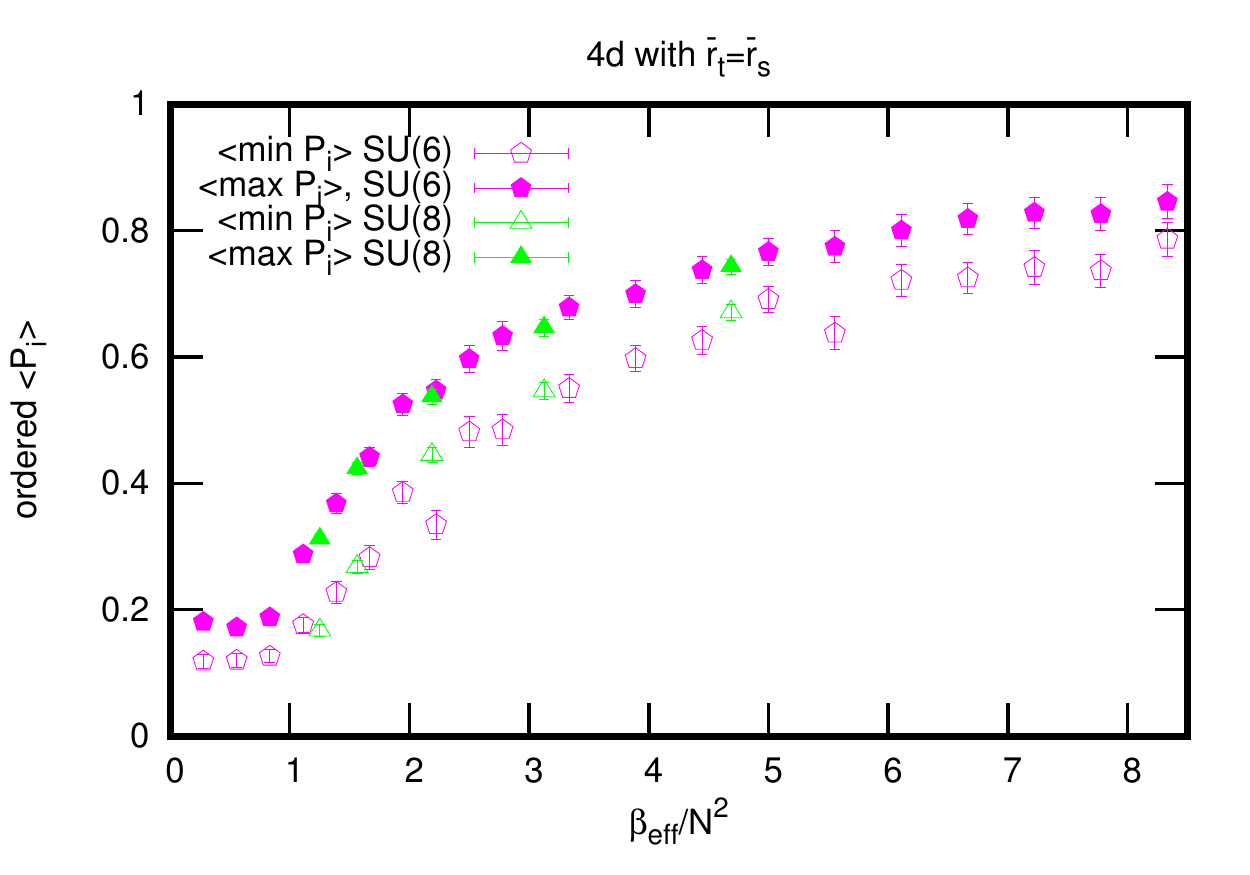}
 \caption{\label{fig:Zn} Expectation value of the ordered Polyakov and Wilson loops for the 2d theory (left) and the 4d theory (right) for equal size lattices. }
  \end{figure}

When the circumferences of the temporal and spatial circles are taken to be the same, $\bar{r}_t=\bar{r}_s$ (for $p=2$), or $\bar{r}_t=\bar{r}_x=\bar{r}_y=\bar{r}_z$ (for $p=4$), and if $\bar{r}_t$ is varied from $\infty$ to 0, then the $(\mathbb{Z}_N)^p$ center symmetry should break down sequentially as \hbox{$(\mathbb{Z}_N)^p\to(\mathbb{Z}_N)^{p-1}\to\ldots\to\{1\}$}. Thus in a single configuration, the center symmetry along the temporal direction may be broken (and as a consequence $P_t\simeq 1$), while the spatial direction may still center symmetry and hence $P_s\simeq 0$. Unfortunately, 
since $\bar{r}_t=\bar{r}_s$, $P_t$ and $P_s$ will flip roles from configuration to configuration in the lattice ensemble\footnote{Note that for large $N$, such `tunneling' will be suppressed.}, so that on average $\langle P_t\rangle = \langle P_s\rangle$ even if one of the center symmetries (but not the other) is broken. An example of this 'two-state' signal is shown in Fig.~\ref{fig:2s}.
Therefore, the expectation value of the Polakov/Wilson loops are not sensitive to this partially broken center symmetric phase. In order to construct an operator that is sensitive, we can (for each configuration) order the expectation values of $P_i$ by size and average the ordered values over ensembles, which ensures that flipped roles of $P_t,P_s$ do not average out. 
In Fig.~\ref{fig:Zn} we study this observable as a function of the temperature (for the 2d theory) or the effective lattice coupling $\beta_{\rm eff}$ (for the 4d theory).

As can be seen from Fig.~\ref{fig:Zn}, there is a region ($\bar{T}\in [0.9,1.1]$ for 2d and --- to a lesser extent --- for 4d and $\beta_{\rm eff}\in[1,5]$) where the average of the maximum Polyakov/Wilson loops does not match the minimum Polyakov/Wilson loops. For instance for 2d and $\bar{T}\simeq 1$, the maximal Polyakov/Wilson loop value of about 0.6 indicates that the corresponding eigenvalue distribution is localized (see e.g. Figures \ref{fig:two},\ref{fig:EV2d}), corresponding to a breaking of the associated $\mathbb{Z}_N$ symmetry. Yet the expectation value of the minimum $P_i$ at this temperature indicates that the corresponding center symmetry remains unbroken. Therefore, the results shown for the 2d theory in Fig.~\ref{fig:Zn} suggest unbroken $(\mathbb{Z}_N)^2$ symmetry for temperatures $\bar{T}\lesssim 0.9$, a remaining $(\mathbb{Z}_N)^1$ symmetry for $\bar{T}\in [0.9,1.1]$ and fully broken center symmetry for $\bar{T}\gtrsim 1.1$. For the 4d theory, the $N=6,8$ results available do not allow to clearly identify a sequential center symmetry breaking pattern. We intend to return to this issue in a future study for larger $N$.


\subsection{$d=10$ $\rightarrow$ 4d}
\label{sec:4d}

\begin{table}[t]
  \centering
\begin{tabular}{|c|cc|c|}
  \hline
  $N$ & $N_t$ & $N_s$ & $\beta_t/N^2$\\
  \hline
  3 & 2 & 8 & 0.546(2) \\
  3 & 3 & 12 & 0.612(1) \\
  3 & 4 & 16 & 0.63(1) \\
  4 & 2 & 8 & 0.591(3) \\
  6 & 2 & 8 & 0.619(3) \\
  \hline
\end{tabular}
\caption{\label{tab:one} Table of critical 4d lattice coupling values for various fixed-size lattices and N.}
\end{table}

\begin{figure}[t]
  \centering
  \includegraphics[width=0.49\textwidth]{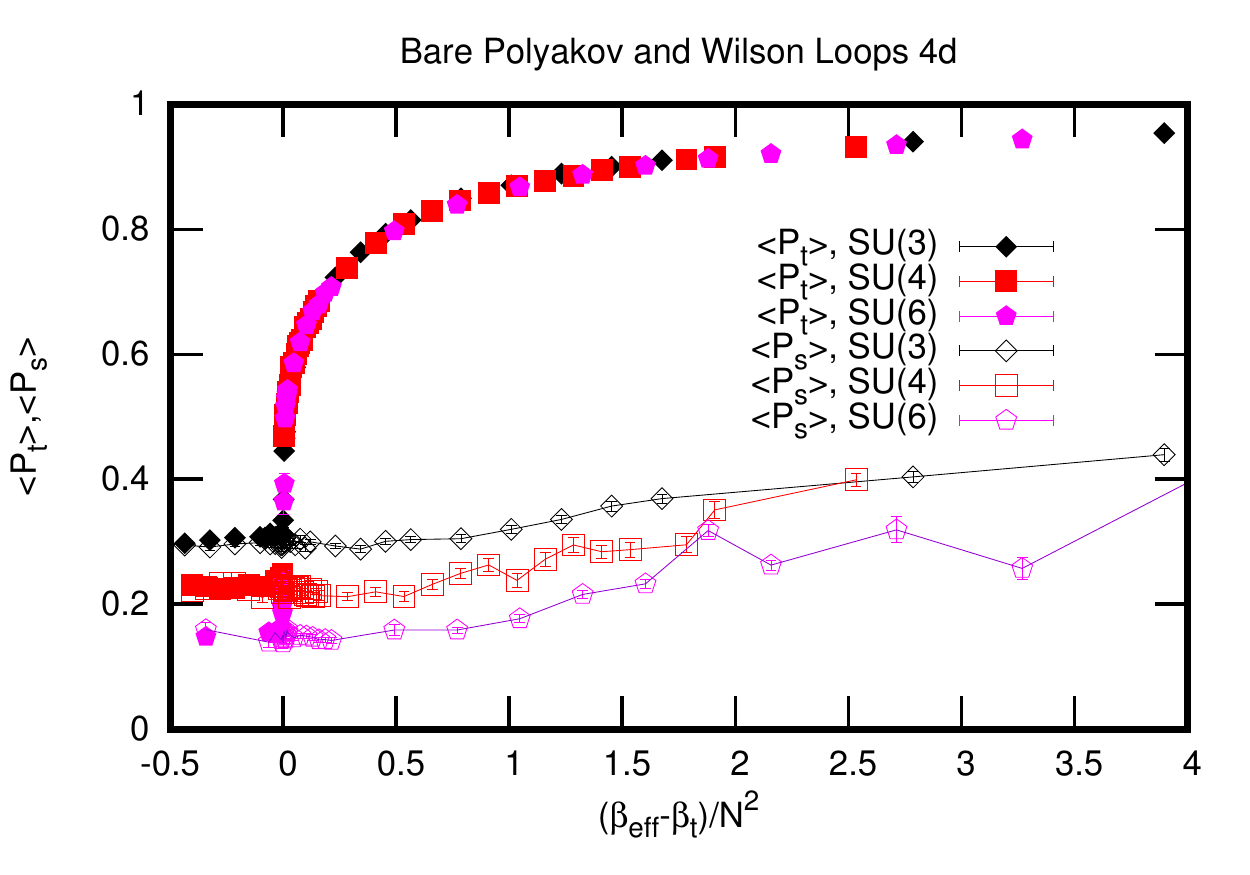}
\hfill
 \includegraphics[width=0.49\textwidth]{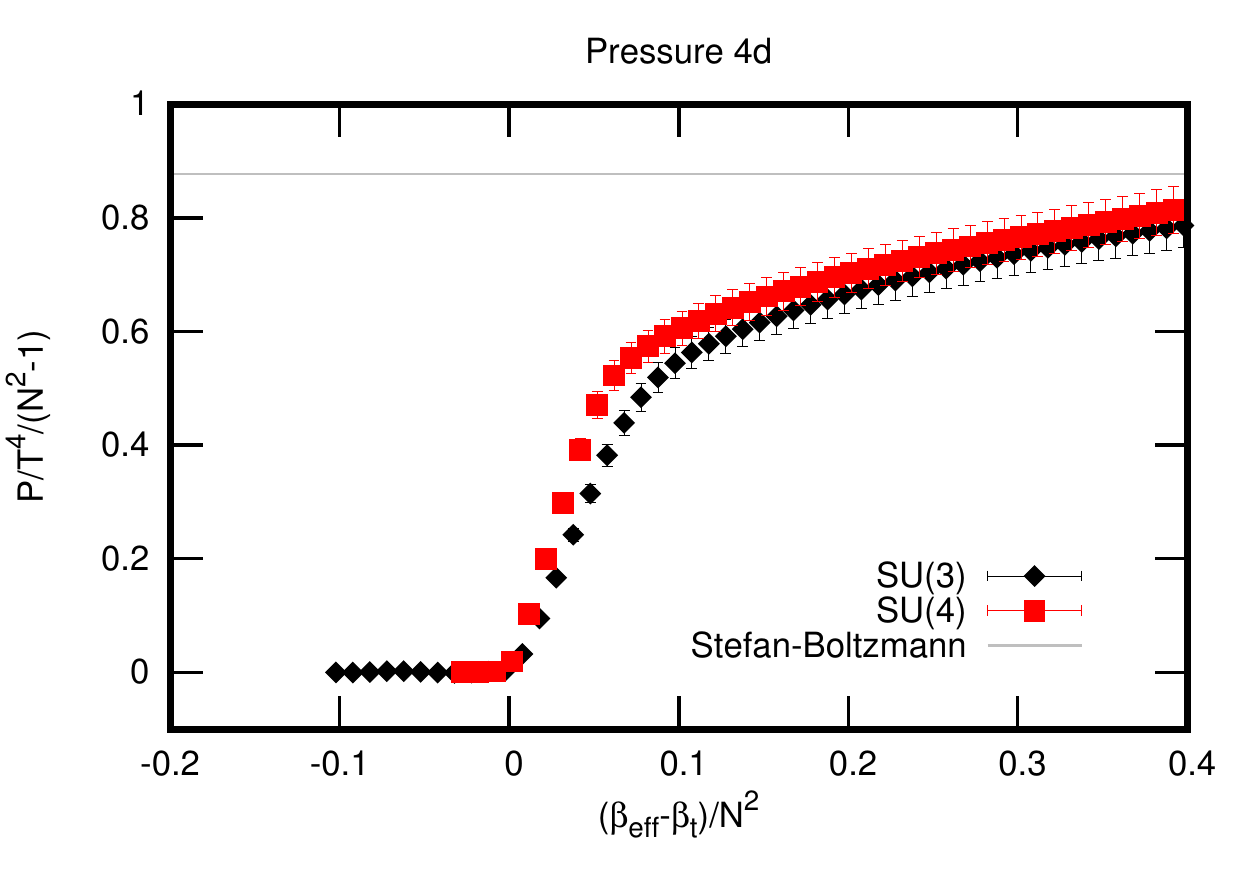}
 \caption{\label{fig:4d} 4d simulations on a $N_t=2$, $N_s=8$ lattice as a function of effective lattice coupling $\beta_{\rm eff}$. Left: bare Polyakov and Wilson loops. Right: Pressure as a function of lattice coupling. $\beta_t$ denotes the critical lattice coupling at which the Polyakov loop susceptibility has a peak (see results in Tab.~\ref{tab:one}). The line labeled 'Stefan-Boltzmann' is the Stefan-Boltzmann result for 8 adjoint bosonic degrees of freedom.}
  \end{figure}

If six of the original ten dimensions are compactified we are studying the bosonic part of ${\cal N}=4$ SYM in $(3+1)$ dimensions. This case is special because for $p=4$ the coupling is dimensionless and we would have to perform scale setting in order to express results in terms of physical quantities. We leave this to future work and report results in terms of the effective four-dimensional lattice coupling $\beta_{\rm eff}$. In this work we only consider bare (unrenormalized) results for the Polyakov/Wilson loop expectation values (see e.g. Ref.~\cite{Dumitru:2003hp} for a discussion on how to extract the renormalized quantities from lattice simulations). At low values of $\beta_{\rm eff}$, both the bare Polyakov and Wilson loop expectation values saturates at the same constant value. For a fixed-size lattice, the expectation value of the Polyakov loop changes discontinuously as $\beta_{\rm eff}$ is changed (see Fig.~\ref{fig:4d}). In analogy to QCD studies in $(3+1)$ dimensions \cite{Boyd:1996bx}, we define this value of the lattice coupling as the critical coupling $\beta_t$. In analogy to our results for 2d above, the expectation value for the Wilson loop starts to change at a generically different value of $\beta_{\rm eff}=\beta_s$ (see Fig.~\ref{fig:4d}). As can be seen from Fig~\ref{fig:4d}, for the values of $N$ and lattice volumes simulated here, the change in the expectation value of the Wilson loop is gradual in $\beta$, unlike the dependence seen in the Polyakov loop. For this reason it is difficult to unambiguously identify a critical value $\beta_s$ given the present statistics. For $N=6$ we find $\beta_s\simeq 2.4(1)$ on a 2$\times$8$^3$ lattice. The values of $\beta_{t}$ depend on the chosen lattice size as well as on $N$ and are shown in Tab.~\ref{tab:one}.
Also shown in Fig.~\ref{fig:4d} are the results for the pressure in the 4d theory. From this figure it can be seen that again the critical Polyakov coupling $\beta_t$ corresponds to the location of deconfinement transition. Note that the pressure results were calculated on a coarse ($2\times 8^3$) lattice, which suffers from significant lattice artifacts at high temperature (cf. the discussion in Ref.~\cite{Boyd:1996bx}), which explains why the numerical results do not seem to converge to the continuum Stefan-Boltzmann behavior.

\begin{figure}[t]
  \includegraphics[width=0.49\textwidth]{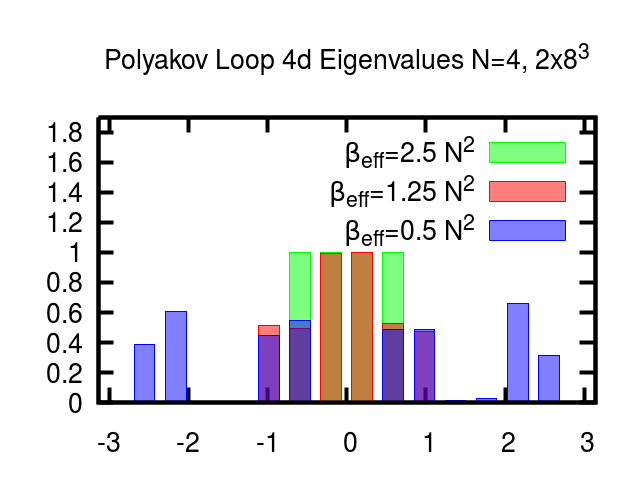}
  \hfill
  \includegraphics[width=0.49\textwidth]{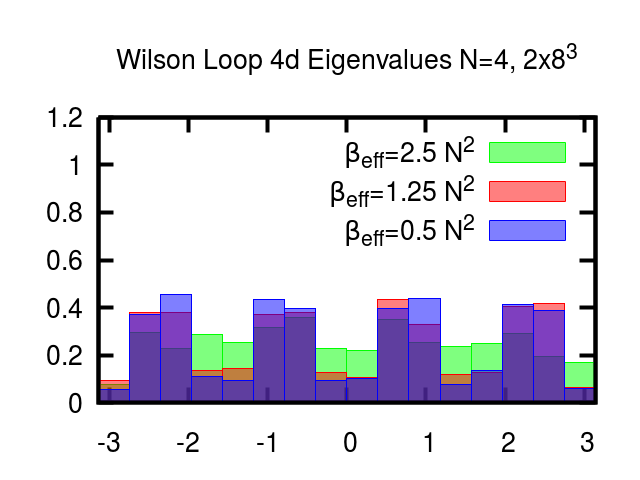}
  \caption{\label{fig:EV4d}  The eigenvalue distribution for the Polyakov loop (left) and Wilson loop (right). Both are for the 4d theory for $N=4$ on a fixed-size lattice of $2\times8^3$ points.  Note the change in the Polyakov loop behavior above/below $\beta_{\rm eff}\simeq 0.56 N^2$ and the Wilson loop behavior above/below $\beta_{\rm eff}\simeq 2.4 N^2$.}
\end{figure}

We proceed to study to eigenvalue distribution for the Wilson and Polyakov loops respectively. In Fig.~\ref{fig:EV4d} the eigenvalue distributions are shown for the case of $N=4$ on a fixed-size lattice of $N_t=2$, $N_s=8$. 
In the left panel, localization of the Polyakov loop eigenvalues can be seen for $\beta_{\rm eff}=1.25$ and $2.5$,
which indicates the system is in the deconfined phase. 
At $\beta_{\rm eff}=0.5$, ${\mathbb Z}_4$-symmetric distribution with four peaks can be seen. 
It is rather different from the $p=1$ and $p=2$ cases, in which rather uniform, continuous distributions have been observed.  
This can be attributed to the larger fluctuations of eigenvalues in lower dimensions due to larger infrared effects. 
A similar situation can be found in the treatment of the eigenvalues of scalar fields in supersymmetric Yang-Mills theories; 
in 3d and 4d one can choose the values by hand because of super-selection, while in 1d and 2d super-selection cannot work and the eigenvalues should be determined dynamically \cite{Banks:1996vh}. 
As for the Wilson loop eigenvalues, four peaks can be seen at $\beta_{\rm eff}=0.5$ and $1.25$, 
while the distribution is uniform at $\beta_{\rm eff}=2.5$. This is natural because  super-selection requires large volume, 
which corresponds to smaller $\beta_{\rm eff}$ for a fixed lattice size. 

To test for the sensitivity to the presence of scalars, we can again change the value of the parent dimension $d$. The $N=3$ results for $\beta_t$ on a $4\times 16^3$ lattice can be readily compared to the case of $d=4$ (pure glue QCD) where $\beta_t(d=4)=5.6908(2)$ \cite{Boyd:1996bx} or $\beta_t(d=4)=0.63231(2)$ (see Ref.~\cite{Panero:2009tv} for a study of pure glue thermodynamics in $d=4$ for different $N$). With our present statistics, this result is numerically indistinguishable from the case of $d=10$, $p=4$ case given in Tab.~\ref{tab:one}. This suggests that the scalars have acquired sizable mass comparable to the temperature scale of the confinement deconfinement transition. In order to simulate smaller scalar masses, we are led to larger lattice spacings (or coarser lattices), e.g. $N_t=2$, $N_s=8$. For this case the we find $\beta_t=0.546(2)N^2$ for $N=3$ for $d=10$, while $\beta_t(d=4)=5.55(11)N^2$, suggesting a mild dependence on the number of scalars. However, the difference between the number of parent theory dimension (which translates to the number of scalars simulated in the 4d theory) is more apparent when considering the action difference $\Delta S$, cf. Eq.~(\ref{eq:deltaS}). In the left panel of Fig.~\ref{fig:10d}, the results for the $\Delta S$ are shown for the theory with 6 scalars and no scalars on a $2\times 8^3$ lattice. Recalling that this quantity effectively is the derivative of the free energy with respect to the lattice coupling (cf. Eq.~\ref{eq:deltaS}), the pronounced difference between the two theories translates into a difference in the free energy at high temperature. Thus, at least on coarse lattices such as $2\times 8^3$, the scalars have not decoupled from the theory.

\subsection{$d=10$}

\begin{figure}[t]
  \includegraphics[width=0.48\textwidth]{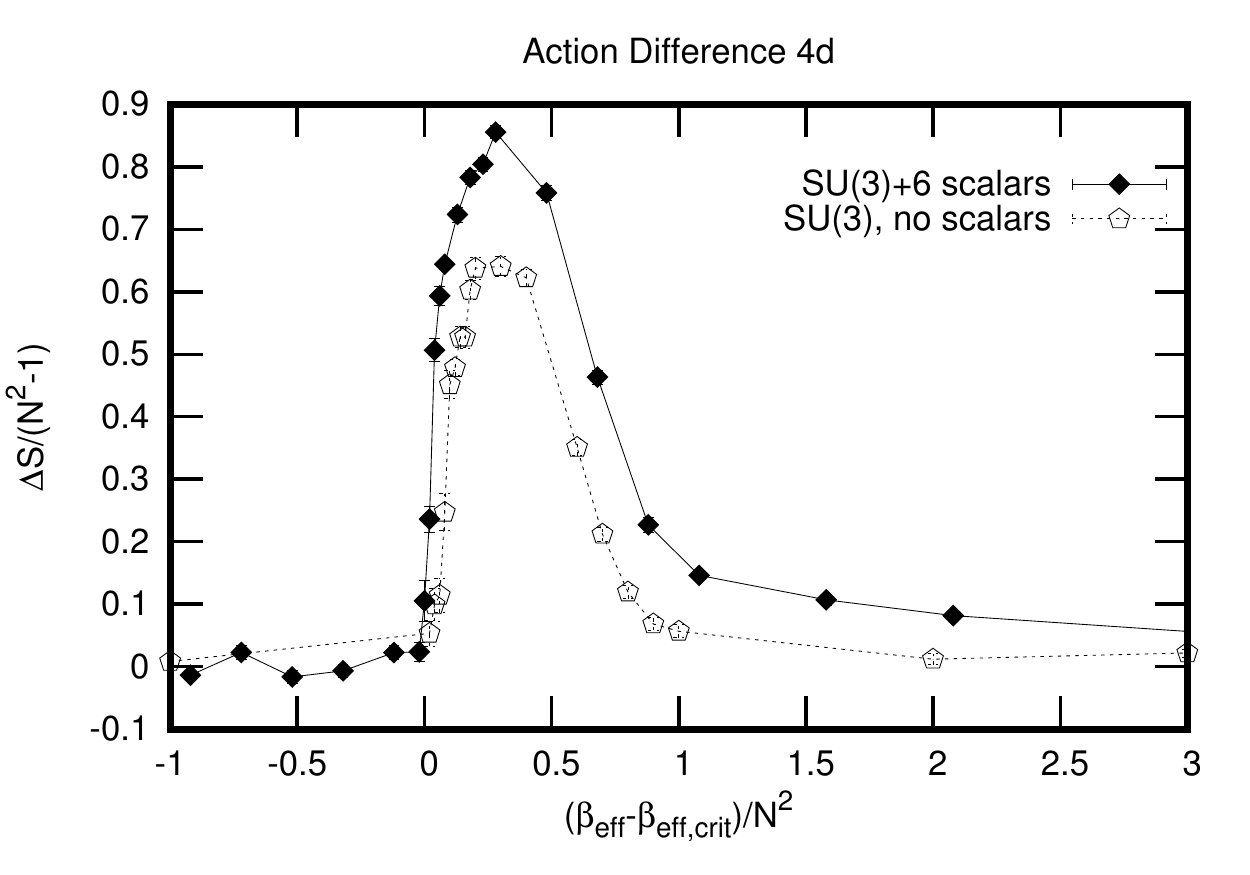}
  \hfill
  \includegraphics[width=0.48\textwidth]{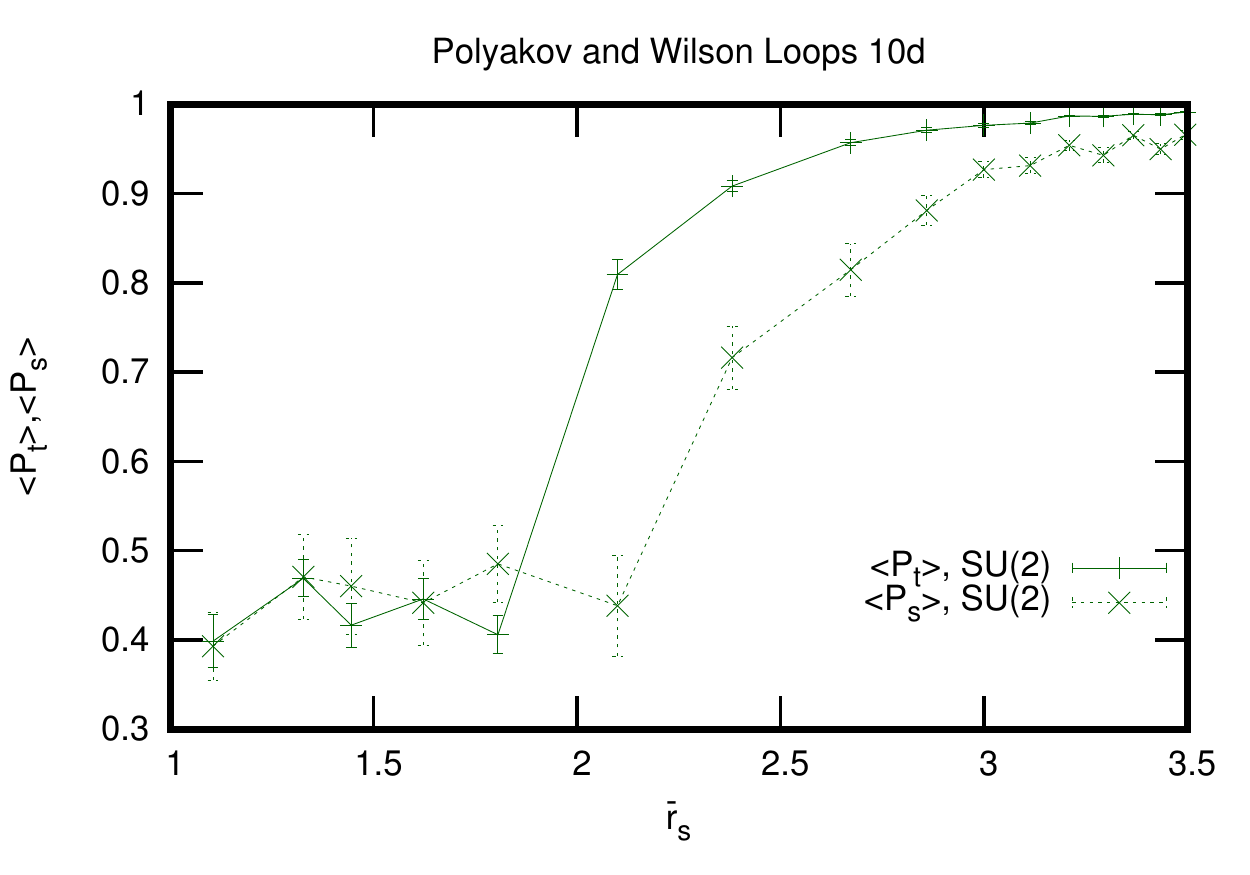}
  \caption{\label{fig:10d}  Left: Simulation results for the action difference $\Delta S$ (cf. Eq.~(\ref{eq:deltaS})) for the 4d theory and N=3 on a $2\times 8^3$ lattice with parent dimensions $d=10$ (``6 scalars'') and $d=4$  (``no scalars''), respectively. Right: $d=10$ simulation results for the Polyakov and Wilson loop expectation values on a fixed size $1\times 3^9$ lattice. }
\end{figure}

It is also technically possible to simulate the original 10d gauge theory on the lattice. This theory is not expected to exist in the sense that it does not have a well defined continuum limit. Yet one may potentially be interested in the 10d theory as a cutoff theory. The lattice spacing $a$ provides such a cutoff and we have performed lattice simulations on a $N_t=1$, $N_s=3$ lattice. The results for the Polyakov and Wilson loop expectation values are shown in the right panel of Fig.~\ref{fig:10d}.

\section{Summary and Conclusions}

Simulating supersymmetric gauge theories on the lattice has interesting connections to gravitational theories via holographic dualities. This work is meant as a step in this direction via simulating the high-temperature phase of these supersymmetric theories, which can be approximated as pure gauge theories in this limit. We have performed lattice simulations of pure gauge 10 dimensional Yang-Mills SU($N$) theory toroidally compactified to $p=1,2,4$ dimensions. The compactification naturally turns the gauge field components in the compactified directions into scalars without any extra programming effort. A possible downside of this approach is the generation of potentially sizable scalar masses. 

We find excellent agreement with simulation results by other group for the 1d case and for the high-temperature limit of the supersymmetric 2d case. We have confirmed the conjectured phase diagram for the bosonic theory for the 2d case and have presented first results for the bosonic 4d case. More work is clearly needed, for instance to determine the scalar masses in the bosonic lattice simulations, so we intend to revisit this (and other) issues in the future. Another future application of this work concerns the real time simulation of the (semi-classical) pure gauge theory dynamics with our code package. The simulation code is flexible concerning the number of parent dimensions $d$ as well as the number of colors $d$, easy to use, and is publicly available as a service to the community \cite{codedown}.

\begin{acknowledgments}

This work was supported, in part, by the Department of Energy, DOE award No. DE-SC0008132. PR would like to thank S. Catterall, A.~Hasenfratz, T.~Ishii, W.~Jay, A.~Kurkela, T.~Morita, E.~Neil, T.~DeGrand and T.~Wiseman for many useful discussions and T.~Wiseman for providing the numerical data obtained in Ref.~\cite{Catterall:2010fx}. This work utilized the Janus supercomputer, which is supported by the National Science Foundation (award number CNS-0821794) and the University of Colorado Boulder. The Janus supercomputer is a joint effort of the University of Colorado Boulder, the University of Colorado Denver and the National Center for Atmospheric Research."

\end{acknowledgments}

\begin{appendix}

\section{Constructing SU($N$) generators}
\label{app1}

To construct the generators $T^a$ for SU($N$) with arbitrary $N$ we employ an algorithm described to us by T. DeGrand. First, let us construct the generators with only off-diagonal entries. A consistent choice is to have off-diagonal generators which are purely real, and purely imaginary. To construct the purely real generators, consider all those $N\times N$ matrices $M_{(1)}$ which have a single entry of ``1'' in the upper triangular part. A purely real generator is constructed by taking $M_{(1)}$ plus its transpose, e.g. $T^{a}=M_{(1)}+M_{(1)}^T$. There are $(N-1)\frac{N}{2}$ of these generators. Next, the purely imaginary generators are constructed by taking all those $N\times N$ matrices which have a single entry of ``i'' in the upper triangular part $M_{(i)}$ plus its hermitian conjugate, e.g. $T^{a}=M_{(1)}+M_{(1)}^\dagger$. There are also $(N-1)\frac{N}{2}$ of these generators.
Finally, there are generators with just entries along the diagonal. These generators are constructed as real diagonal traceless $N\times N$ matrices. Specifically, examples of these are $T^{a}={\rm diag}(1,-1,0,0,\ldots)$, $T^{a}={\rm diag}(1,1,-2,0,0\ldots)$, $T^{a}={\rm diag}(1,1,1,-3,0\ldots)$, etc. There are $N-1$ of these generators.

In total, we have constructed $2\times (N-1)\frac{N}{2}+N-1=N^2-1$ generators for SU($N$). The final step is to normalize the generators such that they obey $Tr \left(T_a T_b\right)=\frac{\delta_{ab}}{2}$. This completes the algorithm to construct the normalized generators for SU($N$).

\section{Comparison between SYM and pure SU($N$) lattice simulations for 2d}
\label{app2}

\begin{figure}[t]
  \centering
  \includegraphics[width=0.6\textwidth]{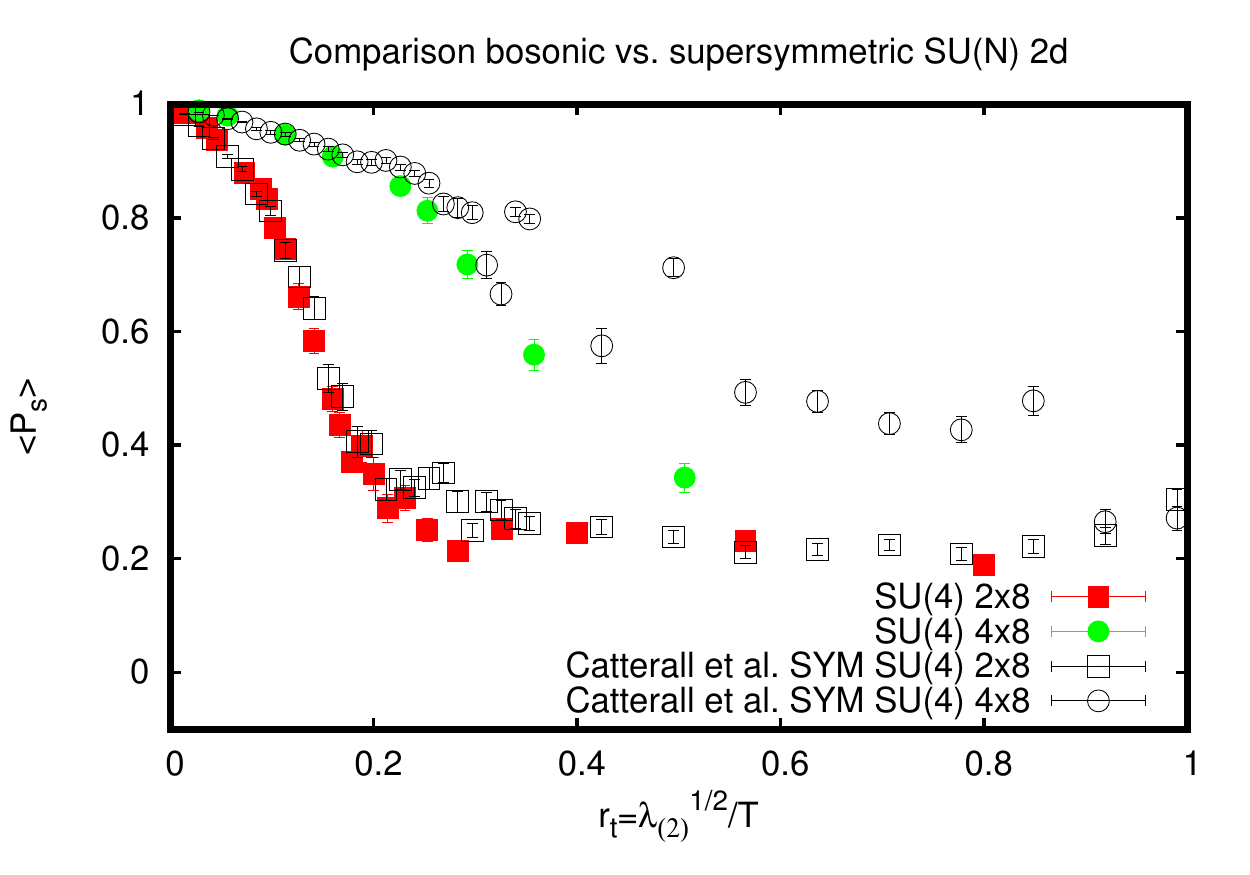}
  \caption{\label{fig:symcomp} Comparison between pure SU($N$) simulations for the 2d case (this work) to lattice simulations of SYM for same-size lattices as a function of temporal radius $\bar{r}_t$. All pure SU($N$) results shown in this figure are for the deconfined phase.}
  \end{figure}

In this appendix we compare the results from our pure SU($N$) simulations in 2d to those for SYM (gauge theory plus fermions) from Ref.~\cite{Catterall:2010fx}. We compare lattice data for the Wilson loop expectation value on a lattices with $N_s=8$ and $N_t=2,4$ for various values of $\beta_{\rm eff}$. Results are shown in Fig.~\ref{fig:symcomp}. Note that results from Ref.~\cite{Catterall:2010fx}, which were calculated using QCD normalization convention \cite{Simonpc}, to holographic normalization convention (cf. Eq.~(\ref{eq:action}).

The comparison shown in Fig.~\ref{fig:symcomp} indicates that results for the SYM simulation indeed coincide with the purely bosonic SU($N$) simulations for high temperatures (low temporal radius). In practice, we find good agreement between the full SYM and bosonic SU($N$) simulations for $\bar{r}_t\leq 0.2$ (corresponding to $T\geq 5 \lambda_{(2)}^{1/2}$).

\end{appendix}

\bibliographystyle{unsrt}
\bibliography{sun}

\end{document}